\documentclass[aps,prl,reprint,superscriptaddress,nolongbibliography]{revtex4-2} %longbibliography
\usepackage{amsmath,amssymb,bm}
\usepackage{graphicx}
\usepackage{dcolumn}
\usepackage{booktabs}
\usepackage{array}
\usepackage{longtable}
\usepackage{hyperref}
\usepackage{xcolor} % In your preamble
\definecolor{revisiongreen}{RGB}{0,120,0}
\newcommand{\rev}[1]{\textcolor{revisiongreen}{#1}}
\hypersetup{hidelinks,hypertexnames=false}

\newcommand{\kvec}{\mathbf{k}}

\newcommand{\rvec}{\mathbf{r}}
\newcommand{\pvec}{\mathbf{p}}
\newcommand{\Sf}{\widetilde{\chi}^{f}_{T}}
\newcommand{\Sv}{\widetilde{\chi}^{v}}
\newcommand{\Sone}{S_{1}}
\newcommand{\Deltau}{\Delta_{u}}
\newcommand{\kcross}{k_{\times}}
\newcommand{\xiscr}{\xi_{\rm scr}}
\newcommand{\rholoc}{\rho_{u}^{\rm loc}}
\newcommand{\Pe}{\mathrm{Pe}}
\newcommand{\dd}{\mathrm{d}}

\makeatletter
\let\arxivsavedmaketitle\maketitle
\let\arxivsavedauthor\author
\let\arxivsavedaffiliation\affiliation
\let\arxivsavedemail\email
\let\arxivsavedthanks\thanks
\let\arxivsavedonecoltitlecolumn\title@column
\long\def\arxivonecoltitlecolumn#1{%
  \minipagefootnote@init
  #1%
  \minipagefootnote@foot}
\makeatother
\begin{document}

% The Letter has no contents page. Suppress only its automatic TOC entries so
% the standard .toc file can remain an independent Supplemental Material TOC.
\makeatletter
\let\arxivsavedaddcontentsline\addcontentsline
\renewcommand{\addcontentsline}[3]{}
\makeatother

%\title{Universal Infrared Leakage in Exchangeable Active Multipoles}

\title{Response-Selected Hidden Hyperuniformity in Hydrodynamic Active Matter}

%\title{Hidden Hyperuniform Flow Order and Its Stability in Active Matter}

\author{Liyu Zhong}
\email{lz8309@princeton.edu}
\affiliation{Department of Mechanical and Aerospace Engineering, Princeton University, Princeton, New Jersey 08544, USA}
\author{Yang Jiao}
\email{yang.jiao.2@asu.edu}
\affiliation{Materials Science and Engineering, Arizona State University, Tempe, Arizona 85287, USA}
\affiliation{Department of Physics, Arizona State University, Tempe, Arizona 85287, USA}

\date{\today}

\begin{abstract}
%\textcolor{blue}{Hyperuniformity in active fluids is usually treated as a property of a prescribed density or continuum field. Here we formulate \emph{response-selected hyperuniformity}, in which long-wavelength order is instead a property of a source--response pair: the incompressible response selects the signed active-moment sector that controls flow. In a reversible valence-one fluid with no prescribed partners, locally neutral clusters generate an analytic $Bk^2$ first-moment spectrum, whereas locally unscreened moments produce a plateau $\Delta_u$. The transverse-force spectrum therefore obeys the universal law $\widetilde{\chi}^{f}_{T}(k)=\Delta_u k^4+Bk^6$, with $k_\times=(\Delta_u/B)^{1/2}$. Complete partner renewal leaves this normal form intact, establishing \emph{exchangeable multipole inheritance}, while turnover tunes $\Delta_u$ through an independently measured local residual density. The zero-residual limit yields strictly hyperuniform velocity fluctuations; any finite residual causes \emph{defect-controlled infrared leakage} and sets a finite screening length. Thus microscopic exchange need not destroy quiet flow, but rare unscreened moments determine how far it survives.}

Hyperuniformity in active matter is usually treated as a property of a prescribed density or continuum field. This view misses a basic feature of hydrodynamic active matter: an incompressible fluid does not respond equally to every microscopic force. Longitudinal forcing is absorbed into pressure, whereas transverse forcing drives flow. The relevant question is therefore not only whether particles are uniformly arranged or whether the total activity is small, but which sector of the active forcing is selected by the physical response. Here we introduce \emph{response-selected hyperuniformity}, in which long-wavelength order is a property of a source--response pair. In a reversible valence-one fluid with no prescribed partners, locally neutral clusters screen the signed active-moment sector that controls transverse flow and generate an analytic \(Bk^{2}\) first-moment spectrum, whereas locally unscreened moments produce a plateau \(\Delta_u\). The transverse-force spectrum obeys the universal law $\widetilde{\chi}^{f}_{T}(k)=\Delta_u k^{4}+Bk^{6}$,
with a crossover \(k_\times=(\Delta_u/B)^{1/2}\). Complete partner renewal leaves this normal form intact, establishing \emph{exchangeable multipole inheritance}, while turnover tunes \(\Delta_u\) through an independently measured local residual density. The zero-residual limit yields strictly hyperuniform velocity fluctuations; any finite residual causes \emph{defect-controlled infrared leakage} and sets a finite screening length. Thus microscopic exchange need not destroy hidden hyperuniform flow order, but rare unscreened moments determine how far the quiet-flow regime survives.
\end{abstract}

\maketitle

%\textcolor{red}{need more background definition of hyperuniformity, in the context of field spectral density and fluctuations.}

\textit{Introduction.---}
Hyperuniform systems strongly suppress fluctuations on wavelengths much larger than
their microscopic scale \cite{Gabrielli2002, TorquatoStillinger2003, Torquato2018}, allowing disordered structures to acquire unusual transport, optical, and mechanical properties
\cite{Donev2005,Zachary2009,Florescu2009,Leseur2016,Chen2021,Jiao2014,Xu17, jiao2026acoustic}. A statistically homogeneous random field \(\phi(\mathbf{x})\) is hyperuniform if its infinite-wavelength fluctuations are suppressed, i.e.,
\begin{equation}
S_\phi(\mathbf{k})
=
\frac{1}{V}\left\langle |\phi(\mathbf{k})|^2\right\rangle
\to 0
\qquad \text{as } |\mathbf{k}|\to 0 .
\end{equation}
Equivalently, let
\begin{equation}
\overline{\phi}_R
=
\frac{1}{v_R}\int_{\Omega_R}\phi(\mathbf{x})\,d\mathbf{x}
\end{equation}
be the field averaged over a window \(\Omega_R\) of linear size \(R\) and volume \(v_R\), then hyperuniformity means
\begin{equation}
\mathrm{Var}(\overline{\phi}_R)
=
\left\langle \overline{\phi}_R^{\,2}\right\rangle
-
\left\langle \overline{\phi}_R\right\rangle^2
=
o(R^{-d})
\qquad \text{as } R\to\infty ,
\end{equation}
i.e., the window-averaged variance decays faster than the \(R^{-d}\) scaling of an uncorrelated field in \(d\) dimensions.

%In active and driven fluids, such suppression has been attributed to random-organization and absorbing-state dynamics, circular motion and effectively center-of-mass-conserving noise, phase-ordering kinetics, and forcing-controlled velocity or vorticity spectra \cite{Corte2008,Hexner2015,Weijs2015,HexnerChaikinLevine2017,Wilken2020, LeiCiamarraNi2019,LeiNi2019,Huang2021,ZhangSnezhko2022,BackofenVoigt2024,ZhengKlattLowen2024,ChenNonreciprocal2024, MaireChaix2025,Maire2026ActiveNucleation}. These studies primarily ask under what dynamics a prescribed density, order parameter, or continuum field acquires vanishing long-wavelength fluctuations.

%Hydrodynamic Active Matter

In hydrodynamic active and driven matter, such suppression has been attributed to several distinct nonequilibrium mechanisms. The earliest and most
influential example is random organization in periodically sheared
suspensions, where particles self-organize toward an absorbing state in
which irreversible collisions are eliminated and large-scale density
fluctuations become anomalously small
\cite{Corte2008,Hexner2015,Weijs2015,HexnerChaikinLevine2017,Wilken2020}.
%This work established the important principle that hyperuniformity need not arise from equilibrium ordering or crystallization, but can emerge dynamically from a disordered driven process.  
Subsequent studies showed that circular or chiral motion, effectively
center-of-mass-conserving active noise, and random-organizing
hydrodynamics can likewise suppress density fluctuations in systems that
remain microscopically disordered
\cite{LeiCiamarraNi2019,LeiNi2019,Huang2021,ZhangSnezhko2022}.
More recently, continuum and field-theoretic approaches have identified
additional routes, including phase-ordering kinetics, conservation-law
constraints, nonreciprocal interactions, and forcing-controlled velocity
or vorticity spectra
\cite{BackofenVoigt2024,ZhengKlattLowen2024,ChenNonreciprocal2024,
MaireChaix2025,Maire2026ActiveNucleation}.
Together, these studies show that nonequilibrium dynamics can generate
vanishing long-wavelength fluctuations in a prescribed density, order parameter, velocity, vorticity, or continuum field.

%In all of these cases, however, the field whose fluctuations are suppressed is also the field being classified as hyperuniform.

%\textcolor{red}{should provide the physical insight here}

Here, we identify a distinctly novel form and mechanism of emergent hyperuniformity in active fluids. The key idea is that the surrounding fluid does not respond equally to every microscopic force in the system. In an incompressible fluid, some force components are absorbed into pressure, while others drive actual flow. So the important question is not simply, “Are the particles arranged uniformly?” or “Is the total activity small?” Instead, the question is: {\it Which part of the active forcing is selected by the physical response?}

Specifically, we define response-selected hyperuniformity as a property of a
source--response pair.  A microscopic source $\widehat q(\kvec)$ generates an
observable $\widehat O(\kvec)=\mathsf G(\kvec)\widehat q(\kvec)$; the response
operator $\mathsf G$ selects the source components that can reach $O$, while
components in its null space are physically silent.  The relevant
long-wavelength order is therefore the spectrum of the selected source sector,
not necessarily that of the total density or unprojected activity. This
perspective is essential in multicomponent and tensorial systems, where the
total density can remain ordinarily fluctuating while a signed source sector
is strongly suppressed \cite{Chremos2017,TorquatoWeighted2026}.  In active
matter, stress, force, pressure, and velocity are tensorial or vectorial fields
\cite{Simha2002,Liverpool2003,Hatwalne2004,Kruse2004,Ramaswamy2010,
Marchetti2013,SaintillanShelley2013,Elgeti2015,Prost2015,Bechinger2016}.
In an incompressible fluid, pressure removes longitudinal force, whereas the
Stokes Green function weights transverse force by $k^{-2}$
\cite{HappelBrenner1983,KimKarrila1991,LaugaPowers2009,KochSubramanian2011}.
The response can therefore select a signed multipole sector whose
long-wavelength order is invisible to number density; we call this
\emph{response-selected hidden hyperuniform flow order}
\cite{SaintillanShelley2007,ThiffeaultChildress2010,Drescher2011,
SpagnolieLauga2012,PushkinYeomans2013,GoldfriendDiamantWitten2017}. Below this definition is implemented literally: the signed first-moment
field is the source, the incompressible transverse projector performs the
selection, and the Stokes Green operator maps the selected force to velocity.

We also investigate whether such order survives continual microscopic rearrangement.
Reversible active assemblies constantly break, diffuse, and rebind
\cite{Sanchez2012,Wensink2012,Takatori2014,Fily2014,Solon2015,Mallory2014}.
We ask whether high-order screening requires permanent molecular identity, and
whether a finite density of locally unscreened moments creates a new exponent
or a crossover. In an exchangeable active fluid with no predetermined
partners, we find that finite neutral clusters generate an analytic $k^2$
first-moment contribution independently of cluster membership, whereas a
generic residual produces an infrared plateau. Complete partner renewal
preserves this multipole inheritance, while the residual produces a universal
$k^6$-to-$k^4$ infrared-leakage crossover and sets the screening length of the hyperuniform flow
regime. These three elements, i.e., response-selected hyperuniformity,
exchangeable multipole inheritance, and defect-controlled infrared
leakage, constitute the organizing framework of this letter.

\begin{figure}[t]
 \centering
 \includegraphics[width=0.48\textwidth]{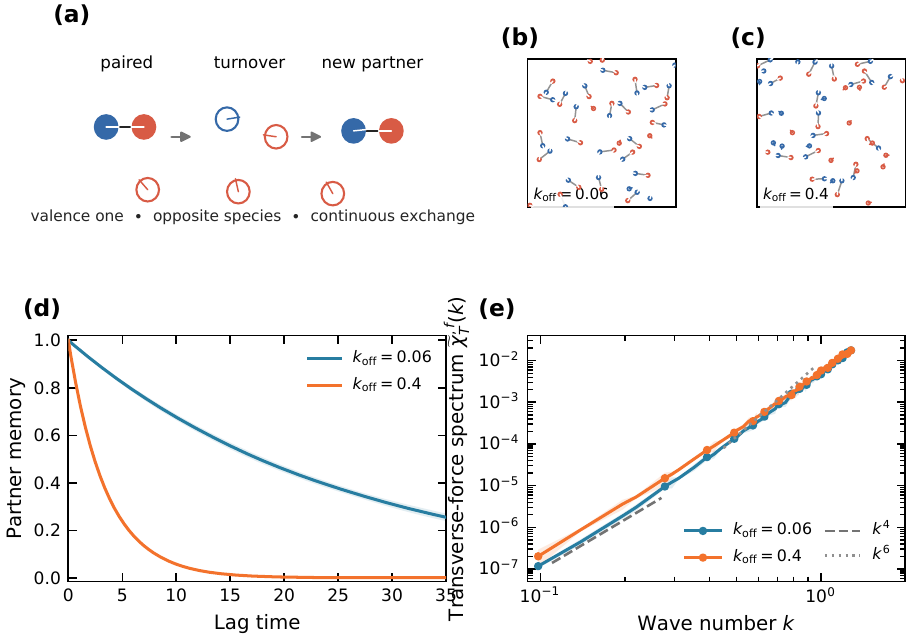}
 \caption{\textbf{Exchangeable active multipoles.}
 (a) Reversible valence-one association with no prescribed partner.
 (b,c) Steady-state crops at $N=1024$ for weak and strong turnover; filled
 particles are associated, open particles are unpaired, and gray segments are
 bonds. (d) Partner memory averaged over four independent seeds. (e)
 Transverse-force spectra; shaded regions are 95\% confidence intervals and
 the gray segments indicate $k^4$ and $k^6$ slopes.}
 \label{fig:exchange}
\end{figure}

\textit{Response-selected hyperuniformity.---}
Let the active units carry signed first moments
$\mathcal M^{(a)}_{ij\ell}=\tau_a\mathcal M_{ij\ell}(\pvec_a)$, where
$\tau_a=\pm1$, see Fig.~\ref{fig:exchange}(a).  Their microscopic source
amplitude is $\widehat{\mathcal A}_{ij\ell}(\kvec)=\sum_a
\mathcal M^{(a)}_{ij\ell}e^{-i\kvec\cdot\rvec_a}$.  With
$\widehat{\kvec}=\kvec/k$ and
$P^T_{ij}=\delta_{ij}-\widehat k_i\widehat k_j$, define
$\mathcal P_T\mathcal M\equiv
P^T_{im}\widehat k_j\widehat k_\ell\mathcal M_{mj\ell}$.  This contraction is
the source sector selected by incompressible flow. Its spectrum is given by
\begin{equation}
 \Sone(k)=\frac{1}{V}\left\langle\left|
 \sum_a \mathcal P_T\mathcal M^{(a)}
 e^{-i\kvec\cdot\rvec_a}\right|^2\right\rangle_{|\kvec|=k}.
 \label{eq:S1}
\end{equation}
Thus $\Sone$ is a source spectrum, while $\mathcal P_T$ is the
response-imposed selection; longitudinal or tensor components annihilated by
this contraction do not contribute to the observed incompressible flow.

Two bound carriers with opposite moments, separation
$\mathbf d$, and center $\mathbf R$ contribute
\begin{align}
 &\mathcal M e^{-i\kvec\cdot(\mathbf R+\mathbf d/2)}
 -\mathcal M e^{-i\kvec\cdot(\mathbf R-\mathbf d/2)}
 \nonumber\\[-2pt]
 &\qquad=-i(\kvec\cdot\mathbf d)\mathcal M e^{-i\kvec\cdot\mathbf R}
 +O(k^2).                                                    \label{eq:pair}
\end{align}
Thus any statistically homogeneous population of finite, locally neutral
clusters contributes $Bk^2+O(k^4)$, irrespective of which microscopic
partners formed each cluster. Here $B\geq0$ is the analytic
neutral-cluster amplitude. Unpaired carriers or imperfectly neutral
clusters instead contribute a nonzero plateau $\Deltau$, which is the variance
density of the response-selected cluster residual. Finite-range correlations
then enforce the source-side analytic normal form, which is derived for arbitrary
finite clusters in SM Sec.~S2.1,

\begin{equation}
 \Sone(k)=\Deltau+Bk^2+O(k^4).                               \label{eq:normal}
\end{equation}

The source-to-force mapping is established as follows: The first-moment
source generates active stress
$\widehat\sigma^a_{ij}=-i\widehat W(k)k_\ell
\widehat{\mathcal A}_{ij\ell}$; taking its divergence and then the transverse
projection gives
$\widehat f_i^T=\widehat W(k)k^2
P^T_{im}\widehat k_j\widehat k_\ell
\widehat{\mathcal A}_{mj\ell}$.  The force amplitude therefore acquires two
powers of $k$, and its spectrum acquires four.  The complete index-level
derivation is given in SM Sec.~S2.1. Consequently, the transverse-force
spectrum is
\begin{equation}
 \begin{aligned}
 \Sf(k)&=\widehat W(k)^2 k^4\Sone(k)\\
 &=\widehat W(k)^2\left(\Deltau k^4+Bk^6+O(k^8)\right).
 \end{aligned}                                                   \label{eq:force}
\end{equation}
where $W$ is a short-range regularization kernel.  The crossover and screening
length are
\begin{equation}
 \kcross=\sqrt{\Deltau/B},\qquad \xiscr=\kcross^{-1},          \label{eq:kx}
\end{equation}
and the two-term running exponent is
\begin{equation}
 \begin{aligned}
 \beta_f^{\rm eff}(k)&\equiv\frac{\dd\ln\Sf}{\dd\ln k}\\
 &=4+\frac{2}{1+(\kcross/k)^2}
 +2\frac{\dd\ln\widehat W}{\dd\ln k}.
 \end{aligned}                                                   \label{eq:beta}
\end{equation}
The last term is a known analytic kernel correction that vanishes as
$k\to0$ (for the Gaussian kernel it is $-2w^2k^2$); it is retained when
predicting finite-window exponents.
For $k\gg\kcross$, locally neutral clusters expose a $k^6$ window; for
$k\ll\kcross$, any finite residual restores $k^4$.  Hence the limits do not
commute:
$\lim_{\Deltau\to0}\lim_{k\to0}\beta_f^{\rm eff}=4$, whereas
$\lim_{k\to0}\lim_{\Deltau\to0}\beta_f^{\rm eff}=6$. This response-selected result uses only local neutrality,
analyticity, and finite correlation length; its leading powers are therefore
independent of a particular reaction scheme. It is response-selected in
a literal sense: $\Sone$ is formed only after the transverse projection, and
the additive $4$ relative to the selected-source running exponent is imposed
by the two-derivative source-to-force map.  A different response operator can select
a different source sector or derivative order from the same microscopic
state.

\begin{figure}[t]
 \centering
 \includegraphics[width=0.48\textwidth]{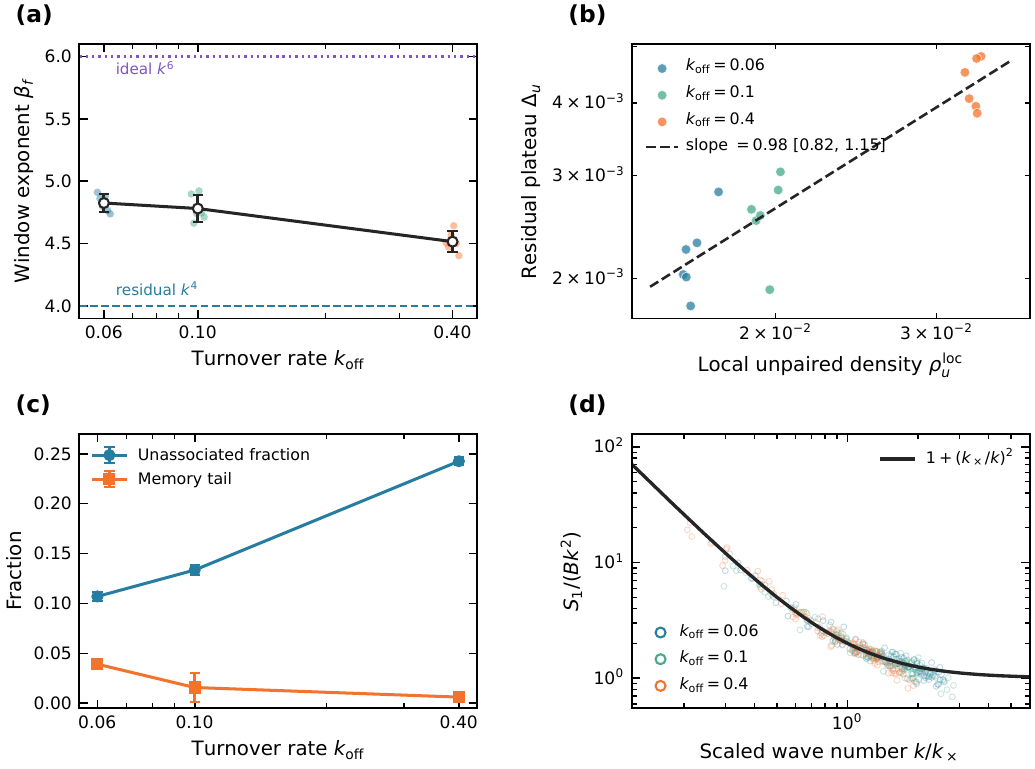}
 \caption{\textbf{Turnover controls a universal infrared crossover.}
 (a) Seed-resolved force-spectrum window exponent; black markers and bars are
 means and 95\% confidence intervals. (b) Spectral plateau versus the
 independently measured local unpaired density.  The dashed line is a joint
 log--log fit, with its bootstrap 95\% interval shown in the legend. (c)
 Unassociated fraction and long-lag memory tail. (d) Collapse of all $N=256$
 spectra onto $\Sone/(Bk^2)=1+(\kcross/k)^2$.}
 \label{fig:turnover}
\end{figure}

\textit{Exchangeable active fluid model.---}
We verify Eqs.~\eqref{eq:normal}--\eqref{eq:beta} in a periodic two-dimensional
fluid of $N$ overdamped active first-moment carriers.  Each carrier has a
position $\rvec_a$, polar axis $\pvec_a=(\cos\theta_a,\sin\theta_a)$, and an
equal-population sign $\tau_a=\pm1$, see Fig.~\ref{fig:exchange}(a). Opposite signs can form a single harmonic
bond if they are sufficiently close and aligned.  Every particle has valence
one, but no partner is assigned in advance.  Bonds form with rate $k_{\rm on}$
and break with a force-accelerated rate whose zero-force value is
$k_{\rm off}$, which governs partner turnover. Translational and rotational Brownian motion, signed active
drift, steric repulsion, and short-range polar alignment act continuously.
Reaction events modify only the bond list: coordinates and orientations are
never reset, and a released carrier may bind any compatible neighbor.  

All
runs start from random unbound configurations. SM Secs.~S3--S5 give
the full stochastic equations, reaction rules, observables, spectral
estimators, parameters, and convergence tests.
Unless stated otherwise, quoted uncertainties are two-sided 95\% Student-$t$
confidence intervals over independent seeds.  We use six seeds at $N=256$ and
four seeds at $N=512$ and 1024.  Stored time samples and wave-vector shells are
averaged within a seed and are never counted as independent realizations.

\textit{Exchangeable multipole inheritance.---}
Figure~\ref{fig:exchange} establishes that the steady state is neither a
quenched molecular crystal nor a fixed-pair construction.  The largest cluster
always contains two particles, but the partner identity decorrelates.  At
$N=1024$, the mean number of distinct partners per particle is $3.98$
$[3.83,4.13]$ for $k_{\rm off}=0.06$ (weak turnover) and $19.05$ $[18.82,19.29]$ for
$k_{\rm off}=0.4$ (strong turnover). The long-lag partner-memory tail is $0.0394$
$[0.0208,0.0580]$ in the former case and $4.02\times10^{-4}$, statistically
consistent with zero, in the latter.  Nevertheless, both states retain a
strongly suppressed transverse-force spectrum [Fig.~\ref{fig:exchange}(e)]. This establishes exchangeable multipole inheritance:
high-order screening is carried by instantaneous local neutrality rather than
by permanent molecular identity.

The crucial effect of partner turnover is not to create a different spectral law, but
to change the balance between neutral and unpaired carriers.  At $N=256$, the
associated fraction decreases from $0.8930$ $[0.8882,0.8978]$ at
$k_{\rm off}=0.06$ to $0.7576$ $[0.7534,0.7617]$ at $k_{\rm off}=0.4$.
Simultaneously, the fitted plateau increases from $2.20$
$[1.83,2.57]\times10^{-3}$ to $4.32$ $[3.88,4.77]\times10^{-3}$, moving
$\kcross$ from $0.548$ $[0.481,0.614]$ to $0.822$ $[0.693,0.951]$.  The ideal
$k^6$ window is consequently pushed to shorter wavelengths.

\textit{Defect-controlled infrared leakage.---}
The force-spectrum exponent measured over the fixed window $k\le0.9$ decreases
continuously from $4.825$ $[4.753,4.897]$ through $4.781$
$[4.671,4.890]$ to $4.515$ $[4.429,4.601]$ as $k_{\rm off}$ increases
[Fig.~\ref{fig:turnover}(a)].  We do not interpret these noninteger values as
new asymptotic exponents: Eq.~\eqref{eq:beta} predicts precisely such values
when the fitting window straddles $\kcross$.

To connect the spectrum to a microscopic observable, we consider an instantaneous cluster as a defect when it has a nonzero
response-selected residual. In the exchangeable valence-one
active-multipole fluid studied here, the dominant explicitly resolved
defect route is an unpaired carrier. We therefore define the
unpaired-moment diagnostic
\begin{equation}
 \rholoc\equiv\frac{1}{V}\left\langle
 \sum_{a\in\mathcal U(t)}
 |\mathcal P_T\mathcal M^{(a)}|^2
 \right\rangle_{t,\widehat{\kvec}},
 \label{eq:rholoc}
\end{equation}
where $\mathcal U(t)$ is the set of unpaired carriers and the angular
normalization is the same as in Eq.~\eqref{eq:S1}; see SM Sec.~S3.3.
This strictly local quantity uses no low-$k$ fit. It does not assume that
every possible imperfect bound-pair residual vanishes; rather, the measured
near-linear relation below shows that unpaired moments dominate the plateau
over the investigated regime.

Across all 18
$N=256$ runs, $\Deltau\propto(\rholoc)^{0.984}$, with bootstrap 95\% interval
$[0.821,1.145]$ and logarithmic Pearson correlation $r=0.929$
[Fig.~\ref{fig:turnover}(b)].  Increasing turnover raises the exchange flux by
more than an order of magnitude, erases partner memory, and increases both
$\rholoc$ and $\Deltau$.  Yet after each seed is scaled by its independently
fitted $B$ and $\kcross$, the complete family collapses onto
$1+(\kcross/k)^2$ given by Eq.~\eqref{eq:beta} [Fig.~\ref{fig:turnover}(d)]. Turnover therefore tunes one infrared-relevant defect
variable rather than changing the universality of the source--response
normal form.

\begin{figure}[t]
 \centering
 \includegraphics[width=0.48\textwidth]{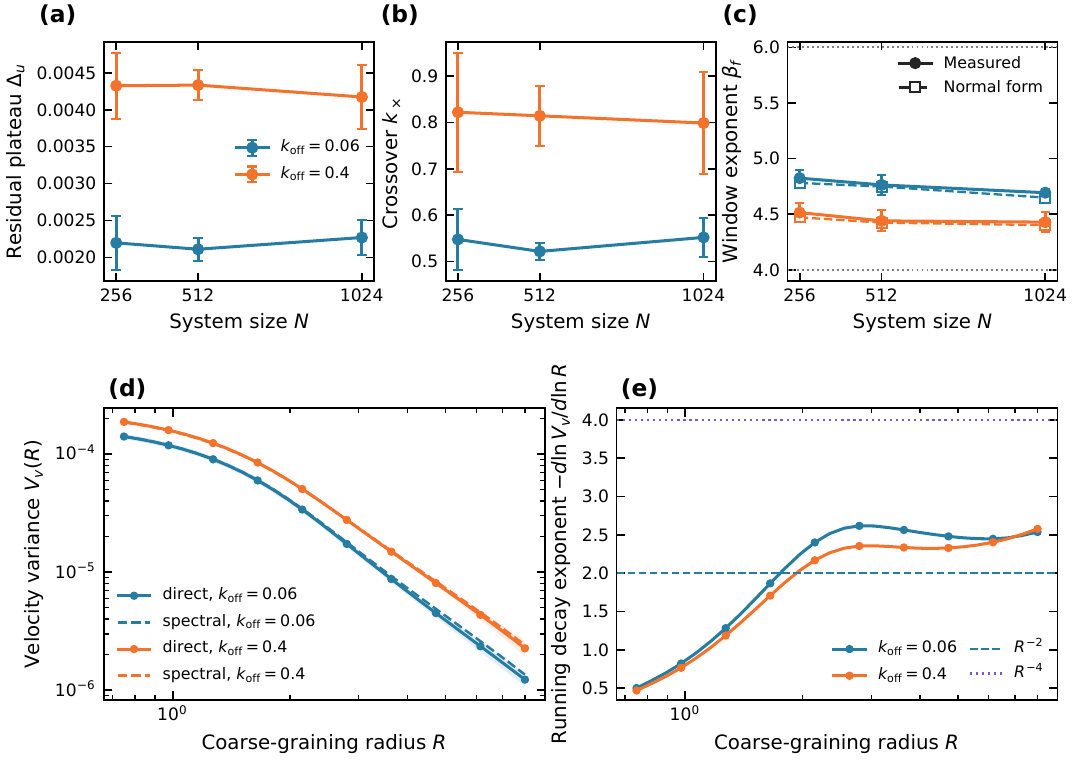}
 \caption{\textbf{Finite-size persistence and macroscopic response.}
 (a,b) Residual plateau and crossover wave number for $N=256$, 512, and 1024.
 (c) Measured window exponent and the value predicted by the two-term normal
 form. Error bars are 95\% confidence intervals. (d) Direct coarse-grained
 velocity variance and its spectral prediction at $N=1024$; shaded regions are
 95\% intervals. (e) Corresponding running decay exponent.}
 \label{fig:sizevelocity}
\end{figure}

\textit{Thermodynamic persistence and velocity leakage.---}
The crossover is not a small-box artifact.  For $k_{\rm off}=0.06$,
$\Deltau$ is $2.20$ $[1.83,2.57]$, $2.11$ $[1.96,2.27]$, and $2.27$
$[2.03,2.51]\times10^{-3}$ at $N=256$, 512, and 1024, respectively.  For
$k_{\rm off}=0.4$, the corresponding values are $4.32$ $[3.88,4.77]$, $4.33$
$[4.13,4.53]$, and $4.17$ $[3.74,4.60]\times10^{-3}$
[Fig.~\ref{fig:sizevelocity}(a)].  The crossover wave numbers are likewise
size independent within uncertainty [Fig.~\ref{fig:sizevelocity}(b)].  Pooling
all 34 runs and sizes gives $\Deltau\propto(\rholoc)^{0.964}$ with bootstrap
95\% interval $[0.884,1.044]$ and logarithmic correlation $r=0.957$.  The
agreement between measured window exponents and Eq.~\eqref{eq:beta} improves
as $k_{\min}/\kcross$ decreases [Fig.~\ref{fig:sizevelocity}(c)].  Thus the
plateau is an intensive defect contribution, while the apparent exponent is a
resolution-dependent crossover observable.

For incompressible Stokes response, the Green operator
$P^T/(\eta k^2)$, with solvent viscosity $\eta$, maps the selected force to
velocity (SM Sec.~S2.2), whose spectrum is given by
\begin{equation}
 \Sv(k)=\frac{\Sf(k)}{\eta^2k^4}\simeq
 \frac{\Deltau+Bk^2}{\eta^2}.                                \label{eq:velocity}
\end{equation}
For the variance $V_v(R)$ of velocity averaged with a Gaussian window
of radius $R$, the spectral integral therefore gives, in two dimensions,
\begin{equation}
 V_v(R)\simeq a\Deltau R^{-2}+bBR^{-4},                       \label{eq:variance}
\end{equation}
with convention-dependent positive constants $a$ and $b$.  We compute the
velocity field directly from 24 steady-state snapshots per seed, average
those snapshots within each seed, and compare its coarse-grained variance
with the prediction obtained from the
measured spectrum.  Their mean ratio remains between $0.905$ and $0.998$ for
$k_{\rm off}=0.06$ and between $0.939$ and $1.010$ for $k_{\rm off}=0.4$
[Fig.~\ref{fig:sizevelocity}(d)].  The running exponent crosses the $R^{-2}$
value and bends away from the ideal $R^{-4}$ decay [Fig.~\ref{fig:sizevelocity}(e)]. The same local residual therefore closes the causal
source--response chain: it controls the force-spectrum leakage, the screening
length, and the restoration of macroscopic velocity noise.

\textit{Discussion and conclusions.---}
The hidden hyperuniform order identified here is conditional, i.e., once the source sector selected
by a physical response organizes into finite, locally neutral clusters, its
long-wavelength limit form is fixed by multipole cancellation and analyticity. The cluster
expansion underlying Eq.~\eqref{eq:pair} (SM Sec.~S2.1) applies to any finite neutral cluster
whose membership can change in time. Cluster chemistry, lifetime, and internal
coordinates determine the amplitudes $\Deltau$ and $B$, but not the two leading
powers or their crossover form. Candidate realizations include reversible
supramolecular assemblies, transiently screened coordination structures, and
defect pairs, provided that their response-selected moments cancel locally.

This mechanism differs from active-fluid hyperuniformity generated by
absorbing-state organization, center-of-mass-conserving dynamics, phase
ordering, or a forcing-controlled turbulent regime. In those settings,
activity reorganizes a prescribed density, scalar order parameter, velocity,
or vorticity field. Here the ordinary particle density need not be
hyperuniform: quietness resides in a signed microscopic source sector selected
by the incompressible response. The advance is therefore not another route to
a vanishing structure factor, but a source-to-response framework that
identifies which hidden sector controls an observable and which local defect
spoils its infrared behavior.

The conclusion is also not restricted to the exchangeable model. A distinct
fixed-partner active-molecule model, with internal reaction states and
independently controlled defect routes, exhibits the same
$\Deltau+Bk^2$ moment normal form and the same $\Deltau k^4+Bk^6$ force
crossover in SM Secs.~S6--S8. A nonzero residual is thus an
infrared-relevant perturbation: however small its amplitude, it ultimately
dominates as $k\to0$, while leaving a potentially broad near-$k^6$ window at
finite observational scales. Strict velocity hyperuniformity occurs in the
zero-residual limit; at finite residual, $\xiscr=\sqrt{B/\Deltau}$ quantifies
the range of the hidden quiet-flow regime.
This distinction may matter in active gels, swimmer suspensions, active
nematics, and dense or phase-separated active fluids
\cite{Giomi2013,DeCamp2015,Doostmohammadi2018,Alert2020,Berthier2014,
FlennerBerthierSzamel2016,BerthierFlennerSzamel2019,
MorseCorwinManning2021,TailleurCates2008,CatesTailleur2015,FodorCates2016,
Tjhung2018,NessCates2020,CatesNardini2025,PadhanVoigt2025,Charbonneau2025ActiveCaging}.
In chiral fluids or odd elastic media, response operators can additionally
mix longitudinal and transverse sectors, creating further routes for hidden
residuals to leak into observable motion
\cite{Banerjee2017,Soni2019,Souslov2019,Scheibner2020,Fruchart2021}.

In summary, we establish a source--response framework for hidden
hyperuniformity in active fluids:
\emph{response-selected hyperuniformity} identifies the hidden source sector
relevant to an observable; \emph{exchangeable multipole inheritance} shows
that its high-order screening survives complete partner renewal; and
\emph{defect-controlled infrared leakage} relates a local residual to the
macroscopic screening length. The broader implication is that hyperuniformity
in multicomponent nonequilibrium systems should not be classified by structure
alone, but relative to an observable and its response operator: different
projections of the same microscopic state can be hyperuniform, ordinarily
fluctuating, or singularly contaminated by the same local defects.
Exchangeable active multipoles provide a minimal realization of this
source--response paradigm.


% -----------------------------------------------------------------------------
% Supplemental Material: restore a standalone-like one-column document inside
% the same PDF, including independent S-numbering, page numbers, TOC and refs.
% -----------------------------------------------------------------------------
\AddToHookNext{shipout/after}{\setcounter{page}{0}}
\onecolumngrid
\clearpage
\makeatletter
\@booleanfalse\twocolumn@sw
\makeatother
\makeatletter
\let\addcontentsline\arxivsavedaddcontentsline
\makeatother
\renewcommand{\rev}[1]{\textcolor{black}{#1}}
\setcounter{figure}{0}
\setcounter{table}{0}
\setcounter{equation}{0}
\setcounter{section}{0}
\setcounter{subsection}{0}
\setcounter{subsubsection}{0}
\renewcommand{\thefigure}{S\arabic{figure}}
\renewcommand{\thetable}{S\arabic{table}}
\renewcommand{\theequation}{S\arabic{equation}}
\renewcommand{\thesection}{S\arabic{section}}
\renewcommand{\thesubsection}{\thesection.\arabic{subsection}}
\renewcommand{\thesubsubsection}{\thesubsection.\arabic{subsubsection}}
\setcounter{secnumdepth}{3}
\setcounter{tocdepth}{2}

\makeatletter
\let\maketitle\arxivsavedmaketitle
\let\author\arxivsavedauthor
\let\affiliation\arxivsavedaffiliation
\let\email\arxivsavedemail
\let\thanks\arxivsavedthanks
\let\title@column\arxivonecoltitlecolumn
\setcounter{affil}{0}
\makeatother
\title{\rev{Supplemental Material for ``Response-Selected Hidden Hyperuniformity in Hydrodynamic Active Matter''}}
\author{Liyu Zhong}
\affiliation{Department of Mechanical and Aerospace Engineering, Princeton University, Princeton, New Jersey 08544, USA}
\author{Yang Jiao}
\affiliation{Materials Science and Engineering, Arizona State University, Tempe, Arizona 85287, USA}
\affiliation{Department of Physics, Arizona State University, Tempe, Arizona 85287, USA}
\date{\today}
\makeatletter
\let\arxivsavedlabel\label
\def\label#1{\arxivsavedlabel{SI.#1}}
\makeatother
\maketitle
\makeatletter
\let\label\arxivsavedlabel
\makeatother

\tableofcontents

\section{Scope and organization}
This Supplemental Material has two purposes.  First, it gives a reproducible
definition and complete validation of the exchangeable valence-one model used
in the main text.  Second, it documents a distinct internally fluctuating
active-molecule model in which complementary partners are assigned in
advance.  The two models differ in degrees of freedom, reaction topology,
initialization, and defect routes.  Their common long-wavelength behavior is
therefore evidence for \rev{the same response-selected source--response normal
form} rather than for one special reaction rule.

Sections~\ref{sec:theory}--\ref{sec:newresults} treat the general theory and
the exchangeable model.  Sections~\ref{sec:legacy-model}--\ref{sec:legacy-results}
give the complete fixed-partner model and its numerical tests.  The final
section states the cross-model conclusion and the boundaries of the claim.

For the exchangeable model, all error bars and shaded bands are two-sided 95\%
Student-$t$ confidence intervals over independent simulation seeds unless a
bootstrap interval is explicitly identified.  Six independent seeds are used
for the $N=256$ turnover series; four seeds are used at $N=512$ and 1024.
Time samples, Fourier shells, and multiple particles within a run are averaged
inside that run and are never treated as independent physical realizations.
For the fixed-partner model, the original seed counts and interval conventions
are stated with each table.

\section{General response-selected moment theory}
\label{sec:theory}

\subsection{\rev{Source, response selection, and cluster expansion}}
\label{sec:source-selection}
\rev{Define the microscopic signed first-moment source}
\begin{equation}
 \color{black}
 \widehat{\mathcal A}_{ij\ell}(\kvec)
 =\sum_a\mathcal M^{(a)}_{ij\ell}e^{-i\kvec\cdot\rvec_a}.
 \label{eq:source-amplitude}
\end{equation}
\rev{For a general linear observable, $\widehat O_\alpha(\kvec)=
\mathsf G_{\alpha,ij\ell}(\kvec)\widehat{\mathcal A}_{ij\ell}(\kvec)$.
We call $\widehat{\mathcal A}$ the source, $\mathsf G$ the response, and the
components of $\widehat{\mathcal A}$ outside the null space of $\mathsf G$ the
response-selected source sector.  Response-selected hyperuniformity means
that this selected sector has a vanishing infrared spectrum, even when the
unprojected source or particle density does not.}

Consider localized active first-moment carriers with tensor amplitudes
$\mathcal M^{(a)}_{ij\ell}$ and a short-range regularization kernel $W$.  For
the first-moment source used here, the active-stress amplitude is
\begin{equation}
 \color{black}
 \widehat\sigma^{a}_{ij}(\kvec)
 =-i\widehat W(k)k_\ell\widehat{\mathcal A}_{ij\ell}(\kvec).
 \label{eq:stress-moment}
\end{equation}
\rev{Let $\widehat{\kvec}=\kvec/k$ and
$P^T_{ij}=\delta_{ij}-\widehat k_i\widehat k_j$.  Taking the stress divergence
and applying the transverse projector selected by incompressibility gives}
\begin{align}
 \color{black}
 \widehat f_i^T(\kvec)
 &=P^T_{im}ik_j\widehat\sigma^a_{mj}(\kvec)\\
 &=\widehat W(k)k^2\widehat{\mathcal A}_{T,i}(\kvec),
 \qquad
 \widehat{\mathcal A}_{T,i}
 =P^T_{im}\widehat k_j\widehat k_\ell
 \widehat{\mathcal A}_{mj\ell}.
 \label{eq:selected-force}
\end{align}
\rev{Equivalently, define
$\mathcal P_T\mathcal M\equiv
P^T_{im}\widehat k_j\widehat k_\ell\mathcal M_{mj\ell}$.  The selected-source
spectrum is}
\begin{equation}
 \Sone(k)=\frac{1}{V}\left\langle\left|
 \sum_a\mathcal P_T\mathcal M^{(a)}e^{-i\kvec\cdot\rvec_a}
 \right|^2\right\rangle_{|\kvec|=k}.
 \label{eq:S1-si}
\end{equation}
\rev{Squaring Eq.~\eqref{eq:selected-force}, shell averaging, and dividing by
$V$ yields the exact first-moment identity}
\begin{equation}
 \color{black}
 \Sf(k)=\widehat W(k)^2k^4\Sone(k).
 \label{eq:force-S1-si}
\end{equation}
\rev{This equation separates the two ingredients: local organization fixes the
infrared power of the selected source $\Sone$, while the response map from a
first moment to transverse force contributes the additional factor $k^4$ to
the spectrum.}

Now partition the carriers into instantaneous finite clusters.  For cluster
$c$ centered at $\mathbf R_c$, expand
\begin{align}
 \sum_{a\in c}\mathcal M^{(a)}e^{-i\kvec\cdot\rvec_a}
 =e^{-i\kvec\cdot\mathbf R_c}\left[
 \mathcal Q_c-i k_m\mathcal D_{c,m}
 -\frac12k_mk_n\mathcal Q_{c,mn}^{(2)}+\cdots\right],
 \label{eq:cluster-expansion}
\end{align}
where $\mathcal Q_c=\sum_{a\in c}\mathcal M^{(a)}$ is the cluster residual.
\rev{The residual relevant to the observable is its selected projection
$\mathcal P_T\mathcal Q_c$.} If the cluster is exactly neutral,
$\mathcal Q_c=0$, and its amplitude begins at $O(k)$.  A homogeneous collection
of finite neutral clusters therefore contributes $Bk^2+O(k^4)$ to $\Sone$.
If a finite density of clusters has \rev{$\mathcal P_T\mathcal Q_c\ne0$}, or
if unclustered carriers remain, the diagonal self-correlation produces a
plateau.  Finite-range cluster correlations can
renormalize its amplitude but cannot change its analyticity.  Hence
\begin{equation}
 \Sone(k)=\Deltau+Bk^2+Ck^4+\cdots,
 \qquad \Deltau\ge0,\quad \rev{B\ge0}.
 \label{eq:normal-si}
\end{equation}
Combining Eqs.~\eqref{eq:force-S1-si} and \eqref{eq:normal-si} gives
\begin{equation}
 \Sf(k)=\widehat W(k)^2\left(\Deltau k^4+Bk^6+Ck^8+\cdots\right).
 \label{eq:force-normal-si}
\end{equation}

The leading two terms define
\begin{equation}
 \kcross=\sqrt{\Deltau/B},\qquad
 \xiscr=\kcross^{-1},
 \label{eq:kx-si}
\end{equation}
and running exponents
\begin{align}
 \beta_1^{\rm eff}(k)&\equiv\frac{\dd\ln\Sone}{\dd\ln k}
 =\frac{2Bk^2}{\Deltau+Bk^2},\\
 \rev{\beta_f^{\rm eff}(k)}&\rev{\equiv\frac{\dd\ln\Sf}{\dd\ln k}
 =4+\beta_1^{\rm eff}(k)
 +2\frac{\dd\ln\widehat W}{\dd\ln k}}\\
 &\rev{=4+\frac{2}{1+(\kcross/k)^2}
 +2\frac{\dd\ln\widehat W}{\dd\ln k}.}
 \label{eq:beta-si}
\end{align}
\rev{The kernel term is analytic and vanishes in the infrared.  For
$\widehat W=e^{-w^2k^2/2}$ it equals $-2w^2k^2$.  We retain this known term in
finite-window predictions; after kernel deconvolution, the universal running
exponent lies exactly between 4 and 6.} Therefore a power-law fit over a finite
interval can return a value near this range even though the analytic
asymptotes are fixed.  We use ``window
exponent'' for such finite-interval estimates and do not interpret them as a
new fractional phase.

\subsection{\rev{Incompressible response and real-space variance}}
\label{sec:stokes-response}
\rev{The Stokes Green operator completes the source--selection--response chain.
Its transverse projector annihilates longitudinal force (which is balanced by
pressure), whereas its factor $k^{-2}$ determines how the selected transverse
force is converted into velocity.} For a two-dimensional incompressible
Stokes response,
\begin{equation}
 \widehat v_i(\kvec)=\frac{1}{\eta k^2}P^T_{ij}(\widehat\kvec)
 \widehat f_j(\kvec),
 \qquad \Sv(k)=\frac{\Sf(k)}{\eta^2k^4}.
 \label{eq:stokes-si}
\end{equation}
At small $k$, $\Sv(k)\simeq(\Deltau+Bk^2)/\eta^2$.  \rev{Thus
$\Deltau=0$ gives a hyperuniform velocity spectrum $\Sv\sim k^2$, whereas any
finite $\Deltau$ restores a velocity plateau in the strict infrared.} For a
Gaussian coarse-graining window $\widehat G_R(k)=e^{-k^2R^2/2}$, \rev{define
$\overline{\mathbf v}_R$ as the window-averaged velocity and
$V_v(R)=\langle|\overline{\mathbf v}_R|^2\rangle-
|\langle\overline{\mathbf v}_R\rangle|^2$.  Then}
\begin{align}
 V_v(R)&=\int\frac{\dd^2k}{(2\pi)^2}\Sv(k)e^{-k^2R^2}\\
 &\simeq a\Deltau R^{-2}+bBR^{-4},
 \label{eq:variance-si}
\end{align}
where $a$ and $b$ depend only on Fourier and window normalization.  The
infrared plateau thus produces an $R^{-2}$ large-window variance, whereas the
neutral-cluster term produces $R^{-4}$ over the screened window.

\section{Model I: exchangeable valence-one active multipoles}
\label{sec:newmodel}

\subsection{Degrees of freedom and stochastic dynamics}
The system contains $N$ carriers in a periodic square of side
$L=(N/\rho)^{1/2}$ at density $\rho=0.25$.  Carrier $a$ has center
$\rvec_a$, orientation $\pvec_a=(\cos\theta_a,\sin\theta_a)$, sign
$\tau_a=\pm1$, and at most one bond.  The signs are globally balanced.
In dimensionless units, the continuous dynamics between reaction updates is
\begin{align}
 \dot{\rvec}_a&=\Pe\,\tau_a\pvec_a-\mu\nabla_aU
 +\sqrt{2D_t}\,\boldsymbol\xi_a,\label{eq:r-dyn}\\
 \dot\theta_a&=-\mu_r\frac{\partial U}{\partial\theta_a}
 +\sqrt{2D_r}\,\eta_a,\label{eq:theta-dyn}
\end{align}
where the noises are independent unit white noises.  The signed propulsion
lets an aligned opposite-sign pair represent a balanced active multipole.
The interaction energy contains harmonic core repulsion, short-range polar
alignment of compatible opposite signs, and harmonic bonds,
\begin{align}
 U=&\sum_{a<b}\frac{\epsilon_{\rm rep}}{2}
 (1-r_{ab}/\sigma)^2\Theta(\sigma-r_{ab})
 -J\sum_{a<b}h(r_{ab})\delta_{\tau_a,-\tau_b}
 \pvec_a\cdot\pvec_b \nonumber\\
 &+\sum_{(a,b)\in\mathcal B}\frac{k_b}{2}(r_{ab}-r_b)^2.
 \label{eq:potential}
\end{align}
Here $h$ is a compact short-range alignment weight with cutoff $r_{\rm al}$,
and $\mathcal B$ is the instantaneous valence-one bond list.  The production
runs use no additional nonbonded attraction ($\epsilon_b=0$), preventing
macroscopic aggregation.  Forces are capped only as a numerical safeguard at
a value far above their steady-state scale.

Each carrier represents a regularized active first moment
\begin{equation}
 \mathcal M^{(a)}_{ij\ell}=\zeta\tau_a\ell\,
 \mathcal T_{ij\ell}(\pvec_a),
 \qquad \widehat W(k)=e^{-w^2k^2/2},
 \label{eq:carrier-moment}
\end{equation}
where $\mathcal T$ is the polar tensor used consistently in the force and
spectral estimators.  Its detailed normalization changes only $B$ and
$\Deltau$, not the powers in Eqs.~\eqref{eq:normal-si} and
\eqref{eq:force-normal-si}.

\subsection{Association, dissociation, and partner exchange}
Reaction attempts occur every $n_r$ integration steps.  Two unbound carriers
are eligible to associate if
\begin{equation}
 \tau_a=-\tau_b,qquad r_{ab}<r_c,qquad
 \pvec_a\cdot\pvec_b>c_c.
\end{equation}
An eligible pair forms a bond with Poisson probability
$1-e^{-k_{\rm on}\Delta t_r}$, subject to the valence-one constraint.  A bond
breaks with a Bell-type force-accelerated rate
\begin{equation}
 k_{\rm off}(F_b)=k_{\rm off}^{0}
 \exp\!\left(\frac{|F_b|}{F_{\rm Bell}}\right),
 \label{eq:bell}
\end{equation}
or immediately if $r_{ab}>r_{\rm break}$.  A reaction changes only the bond
list.  Positions, orientations, and noise histories are not reset, so
dissociated particles diffuse and actively move before rebinding.  Crucially,
there are no partner labels: any compatible neighbor can become the next
partner.  Every production run begins from a random unbound configuration.

\begin{table}[t]
\centering
\caption{Dimensionless parameters for the exchangeable-model production runs.}
\label{tab:new-parameters}
\begin{tabular}{lll}
\toprule
Quantity & Symbol & Value\\
\midrule
Particle number & $N$ & $256,512,1024$\\
Number density & $\rho$ & $0.25$\\
Time step; total steps & $\Delta t$; $n_t$ & $1.5\times10^{-4}$; $6\times10^5$\\
Equilibration steps & $n_{\rm eq}$ & $3\times10^5$\\
Core diameter; stiffness & $\sigma$; $\epsilon_{\rm rep}$ & $1$; $50$\\
Alignment strength; cutoff & $J$; $r_{\rm al}$ & $24$; $2.0$\\
Bond stiffness; rest length & $k_b$; $r_b$ & $100$; $1.35$\\
Capture; break radii & $r_c$; $r_{\rm break}$ & $1.75$; $2.25$\\
Capture alignment threshold & $c_c$ & $0.5$\\
Association rate & $k_{\rm on}$ & $50$\\
Zero-force dissociation rate & $k_{\rm off}^{0}$ & $0.06,0.10,0.40$\\
Bell force scale & $F_{\rm Bell}$ & $25$\\
Reaction stride & $n_r$ & $10$\\
Activity & $\Pe$ & $8$\\
Diffusion constants & $D_t,D_r$ & $1,0.25$\\
Moment length; strength & $\ell,\zeta$ & $0.75,1$\\
Kernel width; viscosity & $w,\eta$ & $0.25,1$\\
Spectral fit window & $k$ & $k\le0.9$\\
\bottomrule
\end{tabular}
\end{table}

\subsection{Dynamical observables}
\label{sec:dynamical-observables}
The associated fraction $\phi_{\rm assoc}$ is the fraction of carriers with a
bond.  Because of valence one, the largest possible cluster contains exactly
two carriers; the observed largest-cluster fraction is $2/N$ in every
production run.  The exchange flux counts completed partner changes per
particle and unit time after equilibration.  For the stored partner series,
the partner-memory function is
\begin{equation}
 C_p(\Delta t)=
 \frac{\left\langle \mathbf 1[b_a(t)=b_a(t+\Delta t)]
 \mathbf 1[b_a(t)\ge0]\right\rangle_{a,t}}
 {\left\langle\mathbf 1[b_a(t)\ge0]\right\rangle_{a,t}},
 \label{eq:memory}
\end{equation}
where $b_a(t)$ is the partner index and $-1$ denotes an unbound carrier.  We
also count the number of distinct nonnegative partner indices visited by each
carrier.

\rev{A defect is defined here by response, not merely by chemistry: an
instantaneous cluster is defective when its selected residual
$\mathcal P_T\mathcal Q_c$ is nonzero.  In the valence-one Model I, unpaired
single carriers provide the dominant directly resolved defect route.}  The
corresponding local unpaired-moment estimator is independent of the infrared
fit and is defined by
\begin{equation}
 \color{black}
 \rholoc=\frac{1}{V}\left\langle
 \sum_{a\in\mathcal U(t)}
 |\mathcal P_T\mathcal M^{(a)}|^2
 \right\rangle_{t,\widehat\kvec},
 \label{eq:rho-local}
\end{equation}
\rev{where $\mathcal U(t)$ is the instantaneous set of unpaired carriers; the
time average is taken over stored steady-state snapshots and the angular
normalization is identical to Eq.~\eqref{eq:S1-si}.  This estimator measures
the diagonal self-correlation of unpaired selected moments.  A perfectly
neutral bound pair contributes only the analytic $Bk^2$ term; any imperfect
bound-pair residual would instead renormalize $\Deltau$ but is not included in
$\rholoc$.  The strong alignment of bonded pairs and the observed near-linear
$\Deltau$--$\rholoc$ relation show that this omitted contribution is subleading
in the production regime.}

\section{Spectral, fitting, and statistical procedures}
\label{sec:newnumerics}

\subsection{Exact spectra}
\label{sec:exact-spectra}
After discarding the first half of each trajectory, up to 32 equally spaced
snapshots are used for exact particle Fourier summation.  For every allowed
periodic wave vector in the analysis range, we evaluate the projected
first-moment amplitude in Eq.~\eqref{eq:S1-si} and the transverse force
amplitude.  Wave vectors with the same $|\kvec|$ are shell averaged, and shell
spectra are averaged over snapshots within a seed.  \rev{The identity
$\Sf=\widehat W^2k^4\Sone$ from Eq.~\eqref{eq:force-S1-si} is also checked
numerically mode by mode.} No FFT deposition or interpolation is used for the
reported force spectra.

The two-term fit minimizes the log-space residual of
$\Sone(k)=\Deltau+Bk^2$ with nonnegative $\Deltau$ and $B$ over $k\le0.9$.
A free power law $Ak^{\beta_f}$ is fitted over the same physical window.
Models are compared using AICc and prediction on held-out wave-number shells.
Because a restricted crossover can mimic a power law, no single AICc result is
treated as decisive.  The main evidence is the joint normal-form collapse,
local--spectral proportionality, finite-size persistence, and velocity
prediction.

\subsection{Confidence intervals and regressions}
For an observable $x_s$ measured in $n$ independent seeds, the plotted
interval is
\begin{equation}
 \overline x\pm t_{0.975,n-1}\frac{s_x}{\sqrt n}.
 \label{eq:t-ci}
\end{equation}
Pointwise spectral and time-series bands apply Eq.~\eqref{eq:t-ci} after each
curve is reduced to one value per seed at each abscissa.  For the relation
$\log\Deltau=a+\gamma\log\rholoc$, the 95\% interval for $\gamma$ is obtained
by 20,000 nonparametric resamples of the independent runs.  The $N=256$
turnover series gives $\gamma=0.984$ $[0.821,1.145]$; pooling all 34 runs gives
$\gamma=0.964$ $[0.884,1.044]$.

\subsection{Velocity postprocessing}
\label{sec:velocity-postprocessing}
For the $N=1024$ trajectories, 24 steady-state snapshots per seed are
deposited on a $256^2$ periodic grid by cloud-in-cell interpolation.  The
known interpolation kernel is deconvolved below $k=0.9$, and the regularized
Stokes response is evaluated with $\eta=1$ and $w=0.25$.  The velocity is then
coarse grained with circular Gaussian windows at 28 radii from $R=0.75$ to 8.
Direct real-space variances are compared with the integral of the measured
velocity spectrum using the identical discrete Fourier convention.

\section{Complete exchangeable-model results}
\label{sec:newresults}

\begin{figure}[t]
 \centering
 \includegraphics[width=0.96\textwidth]{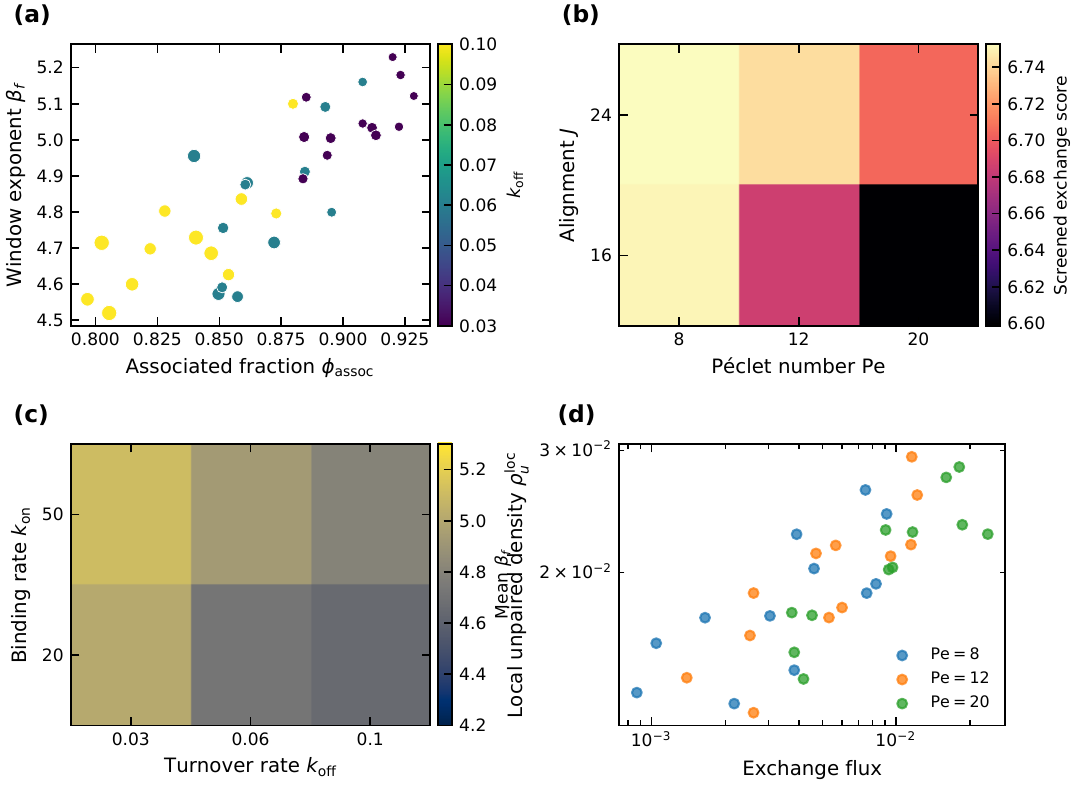}
 \caption{\textbf{Parameter screen for the exchangeable model.}
 (a) Associated fraction and force-spectrum window exponent; marker area
 encodes the number of distinct partners. (b) Maximum screened-exchange score
 over activity and alignment. (c) Mean exponent over association and turnover
 rates. (d) Exchange flux versus local unpaired density.  These pilot runs use
 $N=128$ and identify the nonaggregated, actively exchanging regime used for
 the production simulations.}
 \label{fig:S1}
\end{figure}

\subsection{Parameter selection}
The pilot scans vary $\Pe$, $J$, $k_{\rm on}$, $k_{\rm off}^{0}$, and the
capture-orientation threshold.  Figure~\ref{fig:S1} shows that strong
association alone is not sufficient: vanishing turnover gives persistent
partners, whereas excessive turnover raises the unpaired residual and
contracts the screened window.  The production point $\Pe=8$, $J=24$, and
$k_{\rm on}=50$ combines a high associated fraction with multiple partners
per particle and exactly valence-one clusters.  The main scan then varies only
$k_{\rm off}^{0}$.

\begin{figure}[t]
 \centering
 \includegraphics[width=0.96\textwidth]{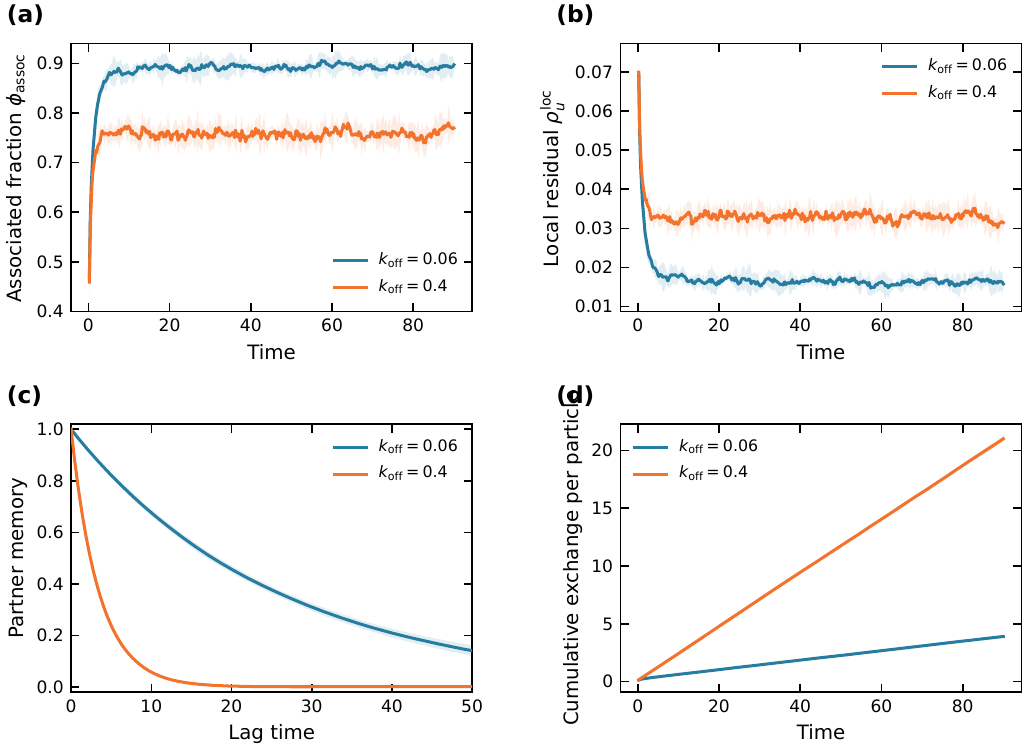}
 \caption{\textbf{Stationarity and exchange at $N=1024$.}
 (a) Associated fraction, (b) local residual, (c) partner memory, and (d)
 cumulative exchange per particle. Curves and shaded bands are independent-seed
 means and 95\% confidence intervals.}
 \label{fig:S2}
\end{figure}

\subsection{Stationarity and partner renewal}
The associated fraction and $\rholoc$ reach stationary plateaus well before the
analysis window begins [Fig.~\ref{fig:S2}(a,b)].  The cumulative reaction count
is linear after this transient [Fig.~\ref{fig:S2}(d)], showing nonzero forward
and reverse activity rather than kinetic arrest.  At strong turnover, the
partner memory is indistinguishable from zero at long lag even though roughly
76\% of particles remain associated.  At weak turnover the memory decays more
slowly, but each particle still samples about four partners on average.

\begin{table}[t]
\centering
\small
\setlength{\tabcolsep}{4pt}
\caption{Turnover dependence at $N=256$. Entries are mean $[95\%\ \mathrm{CI}]$
over six independent seeds.}
\label{tab:turnover-new}
\begin{tabular}{cccccc}
\toprule
$k_{\rm off}^{0}$ & $\phi_{\rm assoc}$ & $\beta_f$ & $\rholoc$ & $\Deltau$ & $\kcross$\\
\midrule
0.06 & $0.8930[0.8882,0.8978]$ & $4.825[4.753,4.897]$ & $0.01628[0.01570,0.01685]$ & $2.20[1.83,2.57]\!\times10^{-3}$ & $0.548[0.481,0.614]$\\
0.10 & $0.8665[0.8617,0.8714]$ & $4.781[4.671,4.890]$ & $0.01953[0.01892,0.02015]$ & $2.58[2.18,2.99]\!\times10^{-3}$ & $0.596[0.512,0.681]$\\
0.40 & $0.7576[0.7534,0.7617]$ & $4.515[4.429,4.601]$ & $0.03297[0.03246,0.03349]$ & $4.32[3.88,4.77]\!\times10^{-3}$ & $0.822[0.693,0.951]$\\
\bottomrule
\end{tabular}
\end{table}

\begin{figure}[t]
 \centering
 \includegraphics[width=0.96\textwidth]{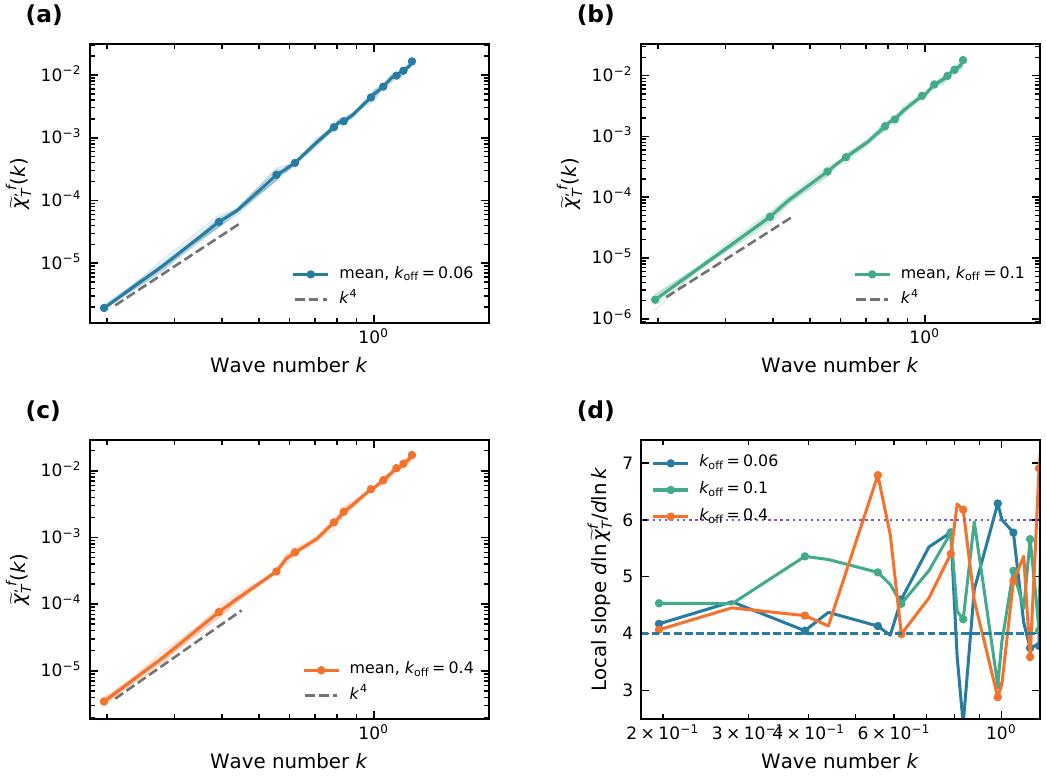}
 \caption{\textbf{Seed-resolved turnover spectra at $N=256$.}
 (a--c) Individual spectra, ensemble means, and 95\% confidence bands.  The
 dashed segments are $k^4$ guides. (d) Local logarithmic slopes of the ensemble
 means; the derivative panel is shown without a confidence band because local
 differentiation strongly amplifies shell noise.}
 \label{fig:S3}
\end{figure}

\subsection{Turnover spectra and finite-window exponents}
Figure~\ref{fig:S3} shows all seed-level spectra behind the main-text turnover
summary.  Increasing $k_{\rm off}^{0}$ raises the low-$k$ amplitude and moves
the crossover to larger $k$.  The local slopes fluctuate because only a small
number of wave-vector shells are available in the deepest infrared.  Stable
claims are therefore based on fixed-window seed fits and the normal-form
collapse, not on a visually selected derivative plateau.  Table~\ref{tab:turnover-new}
lists the corresponding 95\% intervals.

\begin{figure}[t]
 \centering
 \includegraphics[width=0.96\textwidth]{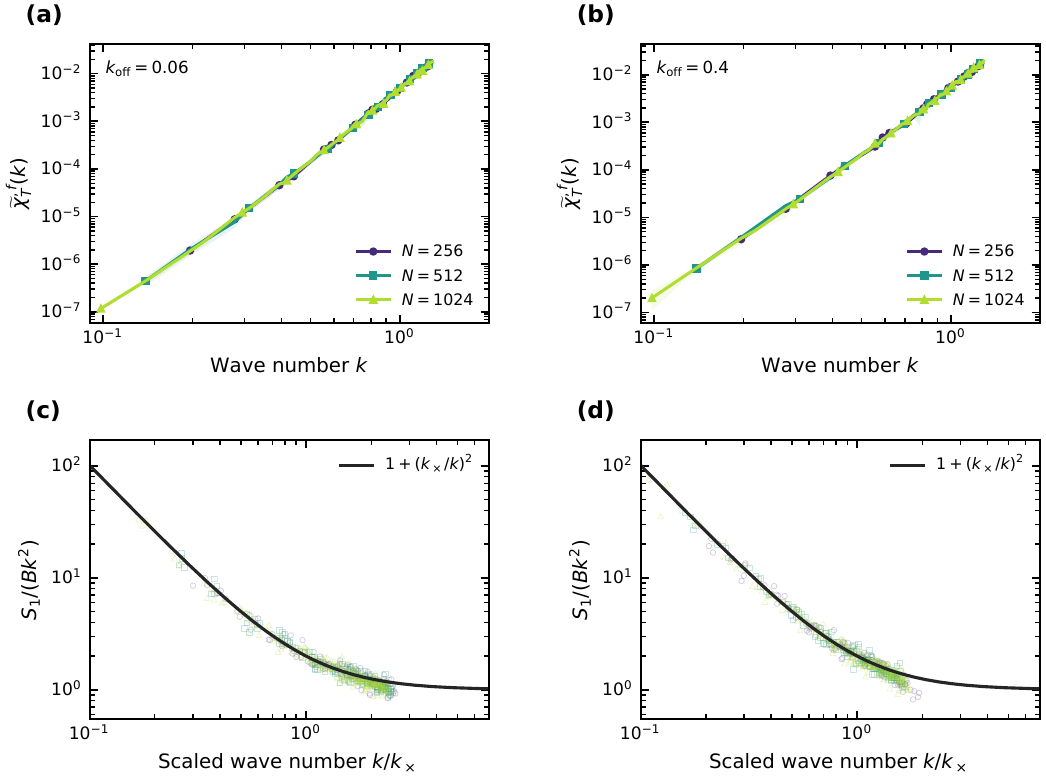}
 \caption{\textbf{Finite-size spectra and normal-form collapse.}
 (a,b) Ensemble transverse-force spectra for $N=256$, 512, and 1024 at weak
 and strong turnover. (c,d) Seed-resolved collapse onto
 $\Sone/(Bk^2)=1+(\kcross/k)^2$.}
 \label{fig:S4}
\end{figure}

\begin{table}[t]
\centering
\caption{Finite-size spectral estimates. Entries are mean $[95\%\ \mathrm{CI}]$;
$n=6$ at $N=256$ and $n=4$ otherwise.}
\label{tab:size-new}
\begin{tabular}{cccccc}
\toprule
$k_{\rm off}^{0}$ & $N$ & $\Deltau$ & $\kcross$ & $\xiscr$ & $\beta_f$\\
\midrule
0.06 & 256 & $2.20[1.83,2.57]\!\times10^{-3}$ & $0.548[0.481,0.614]$ & $1.846[1.635,2.057]$ & $4.825[4.753,4.897]$\\
0.06 & 512 & $2.11[1.96,2.27]\!\times10^{-3}$ & $0.522[0.503,0.541]$ & $1.918[1.848,1.988]$ & $4.764[4.677,4.851]$\\
0.06 & 1024 & $2.27[2.03,2.51]\!\times10^{-3}$ & $0.552[0.510,0.593]$ & $1.815[1.679,1.950]$ & $4.694[4.659,4.728]$\\
0.40 & 256 & $4.32[3.88,4.77]\!\times10^{-3}$ & $0.822[0.693,0.951]$ & $1.240[1.037,1.444]$ & $4.515[4.429,4.601]$\\
0.40 & 512 & $4.33[4.13,4.53]\!\times10^{-3}$ & $0.815[0.750,0.879]$ & $1.230[1.135,1.325]$ & $4.441[4.346,4.536]$\\
0.40 & 1024 & $4.17[3.74,4.60]\!\times10^{-3}$ & $0.799[0.688,0.910]$ & $1.258[1.094,1.423]$ & $4.429[4.341,4.518]$\\
\bottomrule
\end{tabular}
\end{table}

\subsection{Finite-size spectra}
Both turnover limits show stable raw spectra and collapse from $N=256$ to
1024 [Fig.~\ref{fig:S4}].  The minimum resolved wave number relative to the
crossover decreases from mean $k_{\min}/\kcross=0.362$ to $0.178$ at weak
turnover and from $0.244$ to $0.124$ at strong turnover.  As the box grows, the
fit samples more of the plateau-dominated regime; the measured window exponent
therefore drifts toward 4 even though $\Deltau$ and $\kcross$ remain intensive
[Table~\ref{tab:size-new}].  This is the expected signature of a crossover,
not evidence for deteriorating local order.

\begin{figure}[t]
 \centering
 \includegraphics[width=0.96\textwidth]{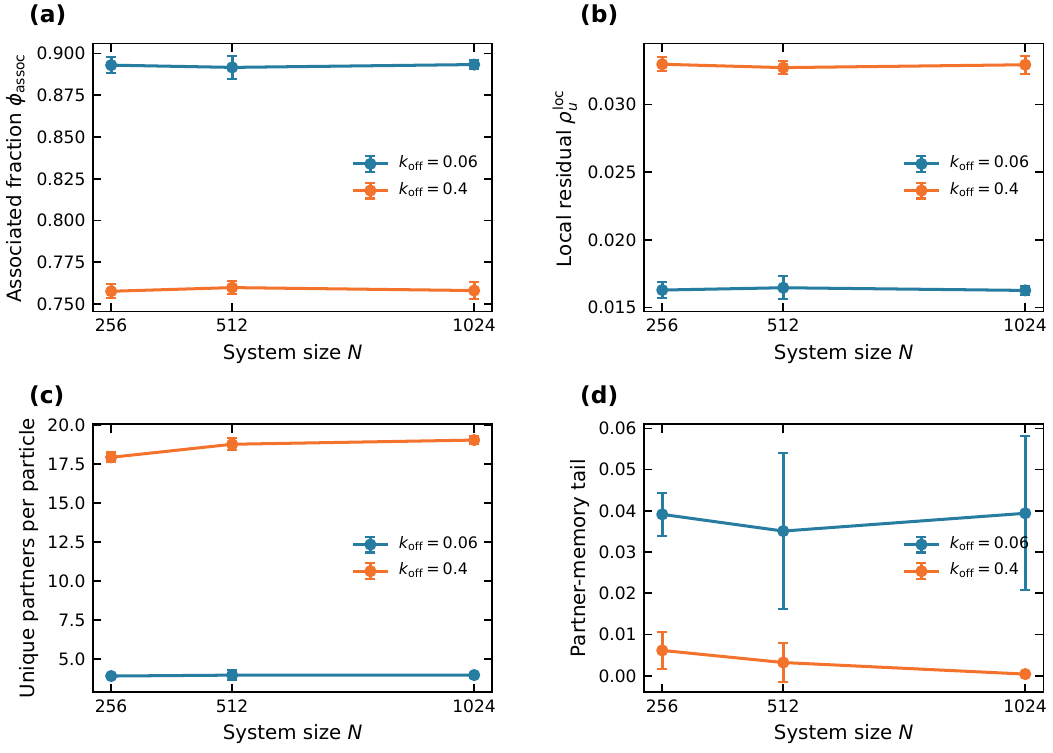}
 \caption{\textbf{Finite-size dynamical observables.}
 Associated fraction, local unpaired density, distinct partners per particle,
 and long-lag partner-memory tail.  Error bars are 95\% seed-level confidence
 intervals.}
 \label{fig:S5}
\end{figure}

\subsection{Finite-size exchange statistics}
The local dynamics are also size independent [Fig.~\ref{fig:S5}].  At weak
turnover, the associated fraction remains near 0.893, $\rholoc$ near 0.0163,
and the number of distinct partners near 4.  At strong turnover the respective
values remain near 0.759, 0.0329, and 19.  The $N=1024$ partner-memory tail is
$0.0394$ $[0.0208,0.0580]$ for $k_{\rm off}^{0}=0.06$ and
$4.02\times10^{-4}$ $[-0.35,8.39]\times10^{-4}$ for $0.4$; the latter interval
is statistically consistent with zero.  These observations rule out a
finite-size locked-partner explanation for the spectral suppression.

\begin{figure}[t]
 \centering
 \includegraphics[width=0.96\textwidth]{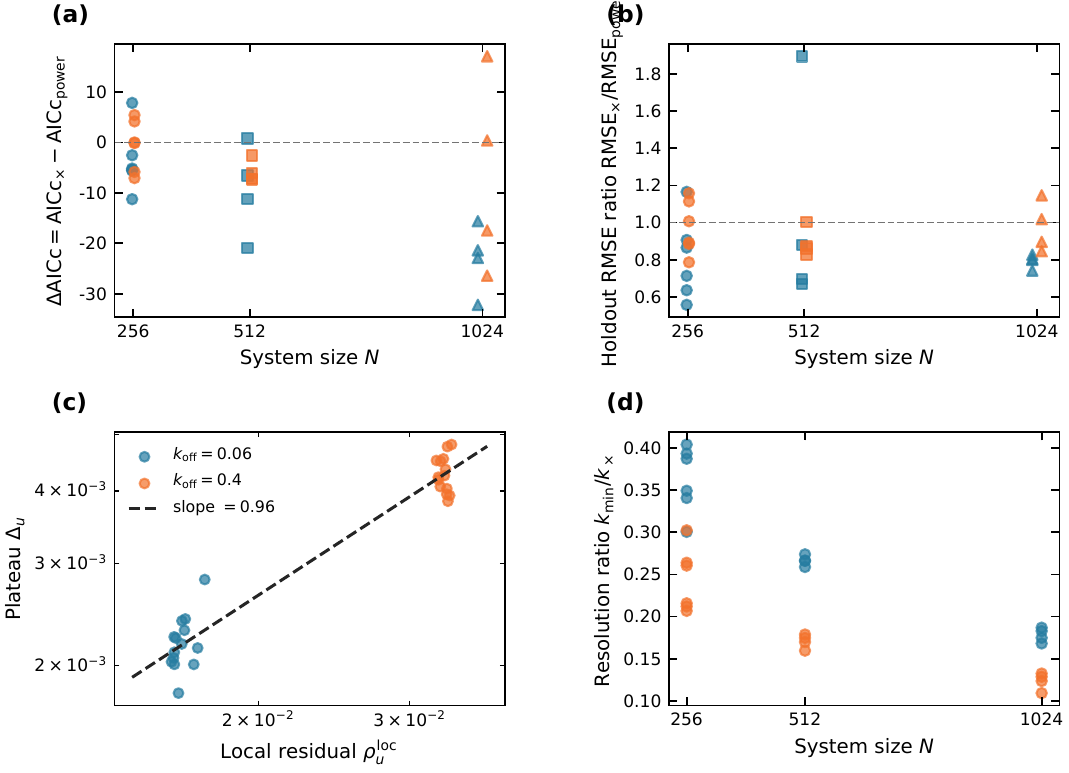}
 \caption{\textbf{Fit and resolution diagnostics.}
 (a) AICc difference between the crossover and free-power models. (b) Ratio of
 held-out-shell errors. (c) Plateau versus local residual for all sizes; the
 dashed line is the joint log--log fit. (d) Resolution ratio
 $k_{\min}/\kcross$.}
 \label{fig:S6}
\end{figure}

\subsection{Model-selection limitations}
The crossover form is not selected by AICc for every individual finite-window
spectrum [Fig.~\ref{fig:S6}(a)].  Depending on size and turnover, a free power
law can approximate the same restricted interval.  Held-out-shell tests show
the same ambiguity [Fig.~\ref{fig:S6}(b)].  The evidence for
Eq.~\eqref{eq:normal-si} is instead the conjunction of: (i) the cluster
analyticity argument; (ii) normal-form collapse across every seed and size;
(iii) an independently measured local residual; (iv) an approximately linear
local--spectral relation with joint slope $0.964$ $[0.884,1.044]$; and (v)
successful prediction of velocity variance.  We therefore make no claim of a
sharp phase transition based on information criteria alone.

\begin{figure}[t]
 \centering
 \includegraphics[width=0.96\textwidth]{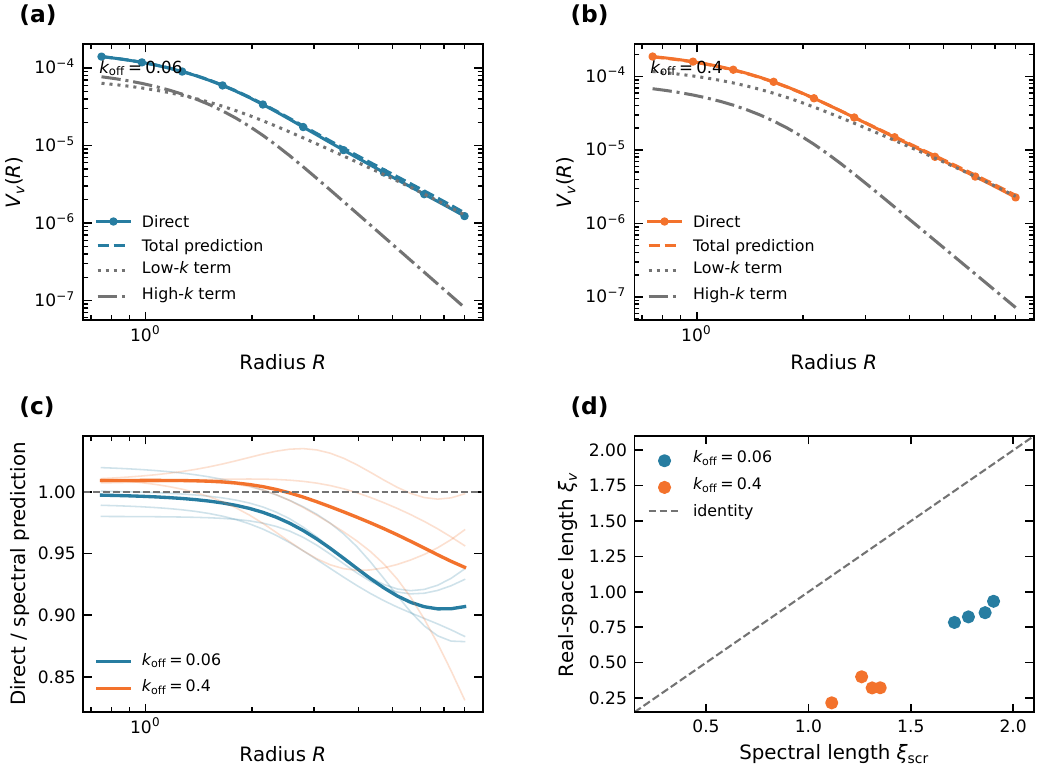}
 \caption{\textbf{Velocity-response diagnostics.}
 (a,b) Direct velocity variances and the total, low-$k$, and high-$k$ spectral
 contributions. (c) Seed-resolved direct/predicted ratios and their means.
 (d) Real-space versus spectral crossover lengths.}
 \label{fig:S7}
\end{figure}

\begin{table}[t]
\centering
\caption{Velocity postprocessing at $N=1024$. Entries are mean
$[95\%\ \mathrm{CI}]$ over four independent seeds.}
\label{tab:velocity-new}
\begin{tabular}{ccccc}
\toprule
$k_{\rm off}^{0}$ & $\xi_{\rm spec}$ & $\xi_v$ & $R_{\times,v}$ & log-fit RMSE\\
\midrule
0.06 & $1.815[1.679,1.950]$ & $0.848[0.747,0.949]$ & $0.810[0.705,0.916]$ & $0.160[0.147,0.173]$\\
0.40 & $1.258[1.094,1.423]$ & $0.315[0.195,0.435]$ & $0.179[-0.028,0.387]$ & $0.116[0.099,0.133]$\\
\bottomrule
\end{tabular}
\end{table}

\subsection{Velocity prediction and length conventions}
The direct velocity variance agrees closely with the integral of the measured
spectrum [Fig.~\ref{fig:S7}(a--c)].  The mean direct/predicted ratio ranges from
0.905 to 0.998 over $R$ at weak turnover and from 0.939 to 1.010 at strong
turnover.  Decomposition into low- and high-$k$ contributions shows how the
plateau term gradually overtakes the analytic $k^2$ term.  The real-space fit
length $\xi_v$ is smaller than $\xiscr$ [Fig.~\ref{fig:S7}(d) and
Table~\ref{tab:velocity-new}] because the Gaussian window, finite kernel,
finite fitting range, and nonasymptotic high-$k$ spectrum alter the numerical
conversion.  We claim a common crossover mechanism, not equality of two
convention-dependent length estimators.

\begin{figure}[t]
 \centering
 \includegraphics[width=0.96\textwidth]{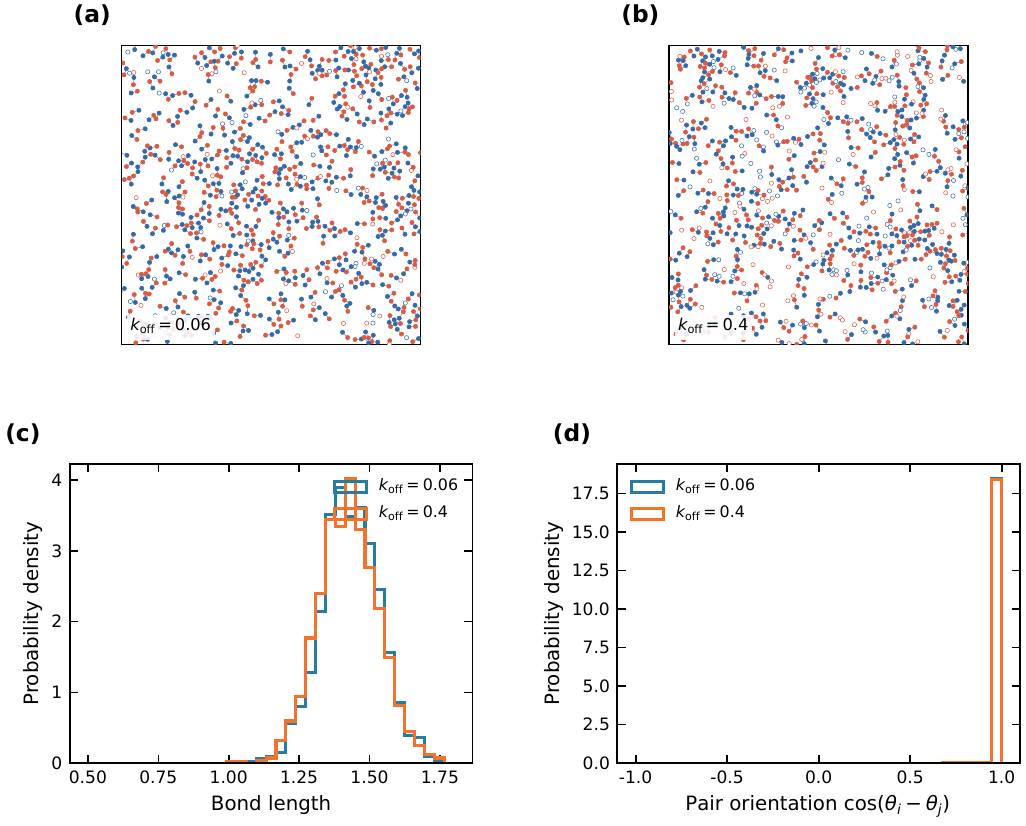}
 \caption{\textbf{Configurations and bonded-pair geometry.}
 (a,b) Full $N=1024$ steady-state configurations at weak and strong turnover.
 (c) Bond-length and (d) pair-orientation distributions pooled across four
 seeds.  The configurations remain homogeneous and nonaggregated.}
 \label{fig:S8}
\end{figure}

\subsection{Geometric checks}
The configurations in Fig.~\ref{fig:S8} show no crystal, phase-separated
domain, or system-spanning cluster.  Bond-length distributions remain centered
near the imposed rest length, and bonded opposite signs are strongly aligned.
Thus turnover mainly changes bond occupancy and partner identity, rather than
creating a different geometric phase.  This separation is important: the
infrared crossover is carried by the signed moment sector, not by ordinary
density ordering.

\section{Model II: internally fluctuating fixed-partner active molecules}
\label{sec:legacy-model}

The second model is retained as an independent realization of the same
infrared principle.  Unlike Model I, it assigns complementary molecular
partners and includes an internal length and three reaction states.  It was
designed to resolve opening, self-healing, escape, wrong-branch selection, and
controlled complementarity perturbations at the event level.  Its preparation
dependence is a limitation of that minimal model, not a feature required by
the general theory.

\subsection{Degrees of freedom}
The simulation contains $N_d$ active dimers in a periodic square of side $L$.
Dimer $a$ has center $\rvec_a$, nematic axis
$\pvec_a=(\cos\theta_a,\sin\theta_a)$, internal length $\ell_a$, signed
first-moment label $\tau_a=\pm1$, and a reaction state.  Complementary dimers
carry partner labels.  A partner pair can be closed, open but locally
associated, or free/escaped.  Chemical state and multipole screening are
distinct: an open pair can remain screened if its signed moments remain
spatially complementary.

\subsection{Continuous dynamics and reactions}
Between reactions,
\begin{align}
 \dot{\rvec}_a &= v_0\tau_a\pvec_a+\mu\mathbf F_a
 +\lambda_H\mathbf u(\rvec_a)
 +\sqrt{2D_t}\,\boldsymbol\xi_a,\\
 \dot\theta_a &= \mu_r T_a+\lambda_H\left[
 \frac12\omega(\rvec_a)+\lambda_J\pvec_a^\perp\cdot
 \mathbf E(\rvec_a)\pvec_a\right]
 +\sqrt{2D_r}\,\eta_a,\\
 \dot\ell_a&=-\mu_\ell\frac{\partial U}{\partial\ell_a}
 +\sqrt{2D_\ell}\,\zeta_a.
 \label{eq:legacy-dynamics}
\end{align}
The interaction contains steric repulsion, finite-depth closed and open Morse
attractions, axis alignment, side matching, length elasticity, and partner
length matching.  Hydrodynamic velocity, strain, and vorticity are evaluated
by a particle--mesh incompressible Stokes solve when $\lambda_H>0$.

The allowed reactions are
\begin{equation}
 \mathrm{closed}\rightleftarrows\mathrm{open},\qquad
 \mathrm{open}\rightleftarrows\mathrm{free}.
\end{equation}
Rates depend on separation, axis mismatch, side mismatch, and active forcing.
A reaction alters only state variables and never resets coordinates,
orientations, or internal lengths.  This model is a minimal active-molecule
normal form motivated by active suspensions, gels, and hierarchical active
assemblies \cite{SI@Ramaswamy2010,SI@Marchetti2013,SI@SaintillanShelley2013,SI@Elgeti2015,SI@Prost2015,SI@Bechinger2016,SI@Sanchez2012}.

\begin{table}[t]
\centering
\caption{Representative parameters for the fixed-partner model.}
\label{tab:legacy-parameters}
\begin{tabular}{lll}
\toprule
Quantity & Symbol & Value or range\\
\midrule
Time step & $\Delta t$ & $5\times10^{-3}$\\
Dimer density & $N_d/L^2$ & $0.32$\\
Translational diffusion & $D_t$ & $0.012$\\
Rotational diffusion & $D_r$ & $0.015$\\
Length diffusion & $D_\ell$ & $0.003$\\
Closed-bond depth & $\epsilon_c$ & $8$\\
Open-bond depth & $\epsilon_o$ & $3.5$\\
Capture radius & $r_c/\ell$ & $1.45$\\
Active speed & $v_0$ & $0.3$--$2.2$\\
Hydrodynamic coupling & $\lambda_H$ & $0$--$1$\\
\bottomrule
\end{tabular}
\end{table}

\subsection{Legacy estimators and confidence intervals}
Exact particle Fourier summation evaluates $\Sone$ and $\Sf$ on a common
low-$k$ vector set.  The local residual is
\begin{equation}
 \rholoc=\frac1V\sum_{c\in\mathrm{pairs}}
 |\mathcal M_{c,+}+\mathcal M_{c,-}|^2
 +\frac1V\sum_{a\in\mathrm{free}}|\mathcal M_a|^2.
 \label{eq:legacy-rho}
\end{equation}
Power-law and crossover fits use fixed physical wave-number windows.  Each
simulation seed is the independent unit; reaction events are summarized
inside a seed before between-seed intervals are formed.  The activity scan
uses stored percentile bootstrap intervals, while reanalyzed observables use
Student-$t$ intervals.  Controlled perturbation replicates are averaged within
their common base seed.

\section{Complete fixed-partner-model results}
\label{sec:legacy-results}

\begin{figure}[t]
 \centering
 \includegraphics[width=0.62\textwidth]{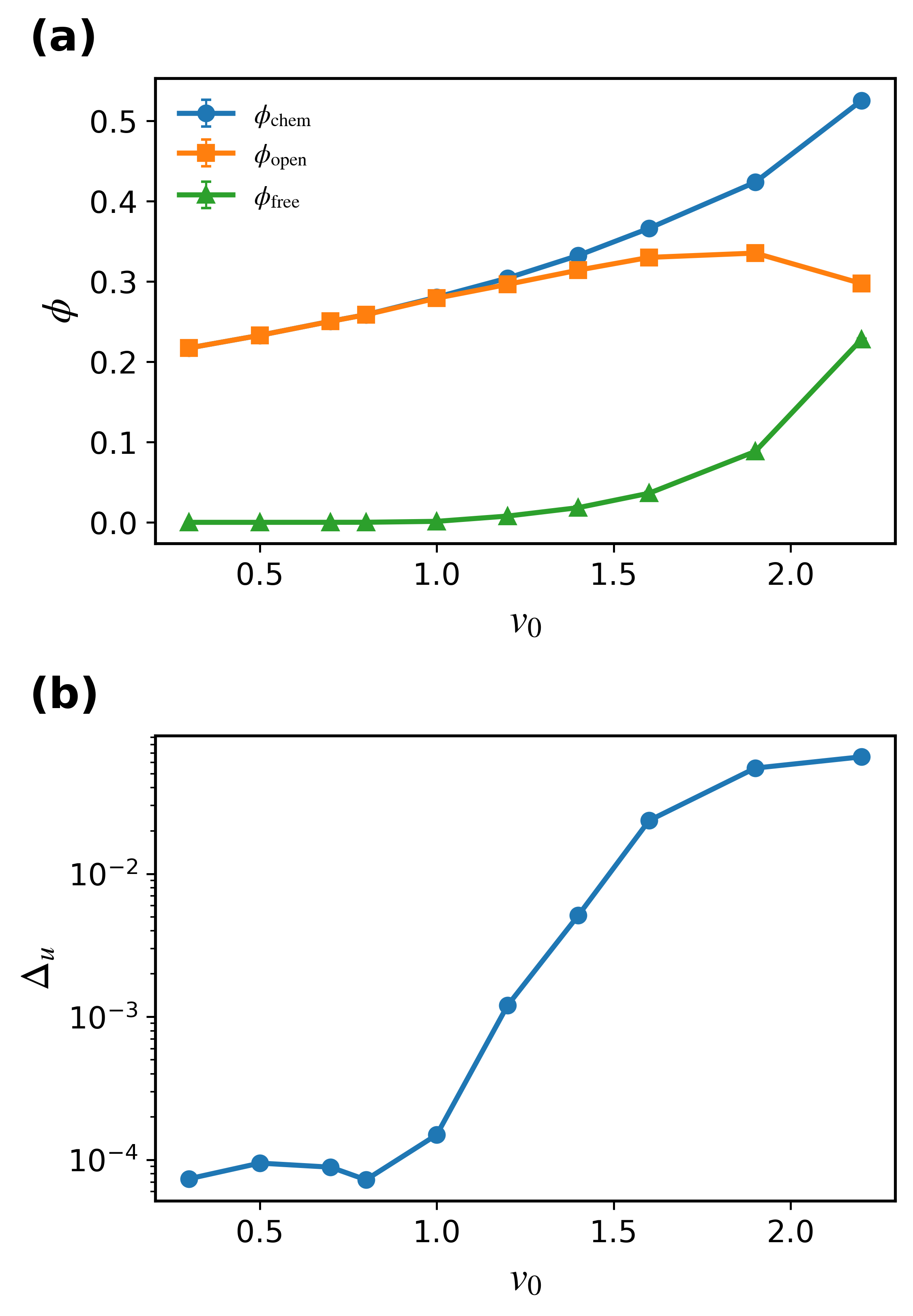}
 \caption{\textbf{Fixed-partner model: activity-driven loss of screening.}
 Steady-state chemical, open, and escaped fractions and the fitted plateau
 $\Deltau$.  Error bars in the fraction panel are 95\% seed-level intervals;
 numerical intervals for the plateau are listed in Table~\ref{tab:legacy-activity}.}
 \label{fig:S9}
\end{figure}

\subsection{Activity-driven self-healing and escape}
At $v_0=0.8$, the chemical-defect fraction is $0.2586$
$[0.2566,0.2610]$ but the escaped fraction is only
$1.04\times10^{-4}$ $[0.92,1.16]\times10^{-4}$.  The plateau is
$7.26\times10^{-5}$ $[1.95,12.56]\times10^{-5}$ and the force exponent is
$5.63$ $[5.45,5.81]$.  At $v_0=1.9$, the escaped fraction rises to $0.0883$
$[0.0870,0.0895]$, the plateau to $0.0549$ $[0.0502,0.0596]$, and the exponent
becomes $3.99$ $[3.94,4.05]$ [Fig.~\ref{fig:S9}].  Chemical opening alone is
therefore not the relevant defect variable; spatially unscreened first moments
are.

\begin{table}[t]
\centering
\caption{Selected fixed-partner activity-scan estimates. Mean
$[95\%\ \mathrm{CI}]$ over eight seeds.}
\label{tab:legacy-activity}
\begin{tabular}{ccccc}
\toprule
$v_0$ & $\phi_{\rm chem}$ & $\phi_{\rm free}$ & $\beta_f$ & $\Deltau$\\
\midrule
0.3 & $0.2172[0.2162,0.2181]$ & $8.11[4.06,12.98]\!\times10^{-6}$ & $5.65[5.48,5.83]$ & $7.38[2.92,11.83]\!\times10^{-5}$\\
0.8 & $0.2586[0.2566,0.2610]$ & $1.04[0.92,1.16]\!\times10^{-4}$ & $5.63[5.45,5.81]$ & $7.26[1.95,12.56]\!\times10^{-5}$\\
1.9 & $0.4236[0.4218,0.4251]$ & $0.0883[0.0870,0.0895]$ & $3.99[3.94,4.05]$ & $0.0549[0.0502,0.0596]$\\
2.2 & $0.5252[0.5237,0.5267]$ & $0.2278[0.2267,0.2290]$ & $3.91[3.86,3.96]$ & $0.0658[0.0625,0.0690]$\\
\bottomrule
\end{tabular}
\end{table}

\begin{figure}[t]
 \centering
 \includegraphics[width=0.96\textwidth]{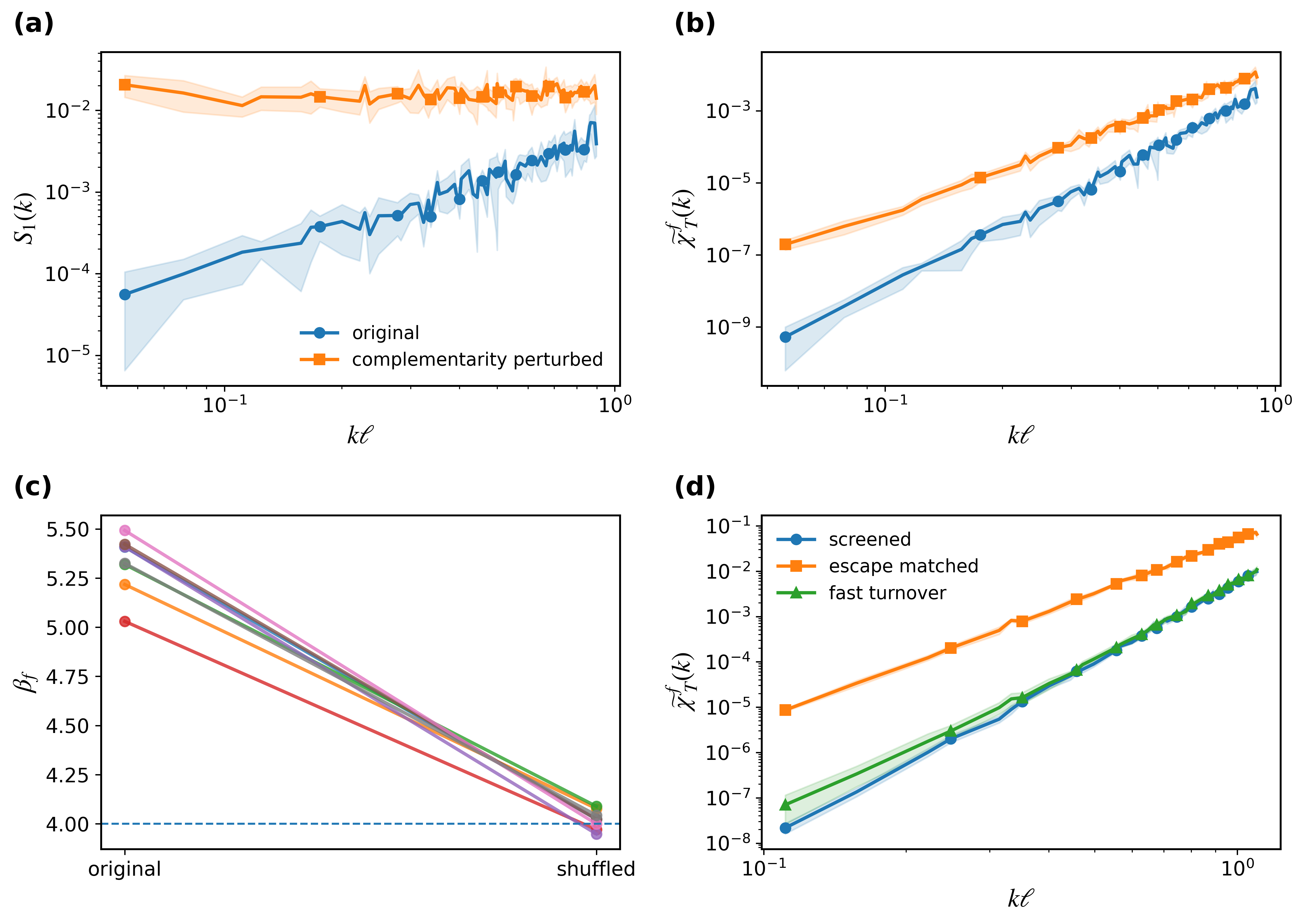}
 \caption{\textbf{Fixed-partner model: complementarity causality test.}
 (a,b) First-moment and force spectra before and after signed-moment
 reassignment at fixed geometry and chemistry. (c) Base-seed-paired exponent
 changes. (d) Matched-chemistry protocols. Bands and bars are 95\%
 seed-level intervals.}
 \label{fig:S10}
\end{figure}

\subsection{Causal complementarity perturbation}
Randomly reassigning signed first moments among fixed centers preserves
positions, axes, lengths, bond states, and one-point moment statistics while
destroying spatial complementarity.  The first-moment spectrum immediately
develops a plateau and the force spectrum moves toward $k^4$
[Fig.~\ref{fig:S10}(a,b)].  Across every base seed, the exponent decreases from
$5.328$ $[5.205,5.450]$ to $4.022$ $[3.981,4.062]$, a paired reduction of
$1.306$ $[1.176,1.436]$ [Fig.~\ref{fig:S10}(c)].  This isolates signed-moment
complementarity from ordinary geometry.

\begin{figure}[t]
 \centering
 \includegraphics[width=0.96\textwidth]{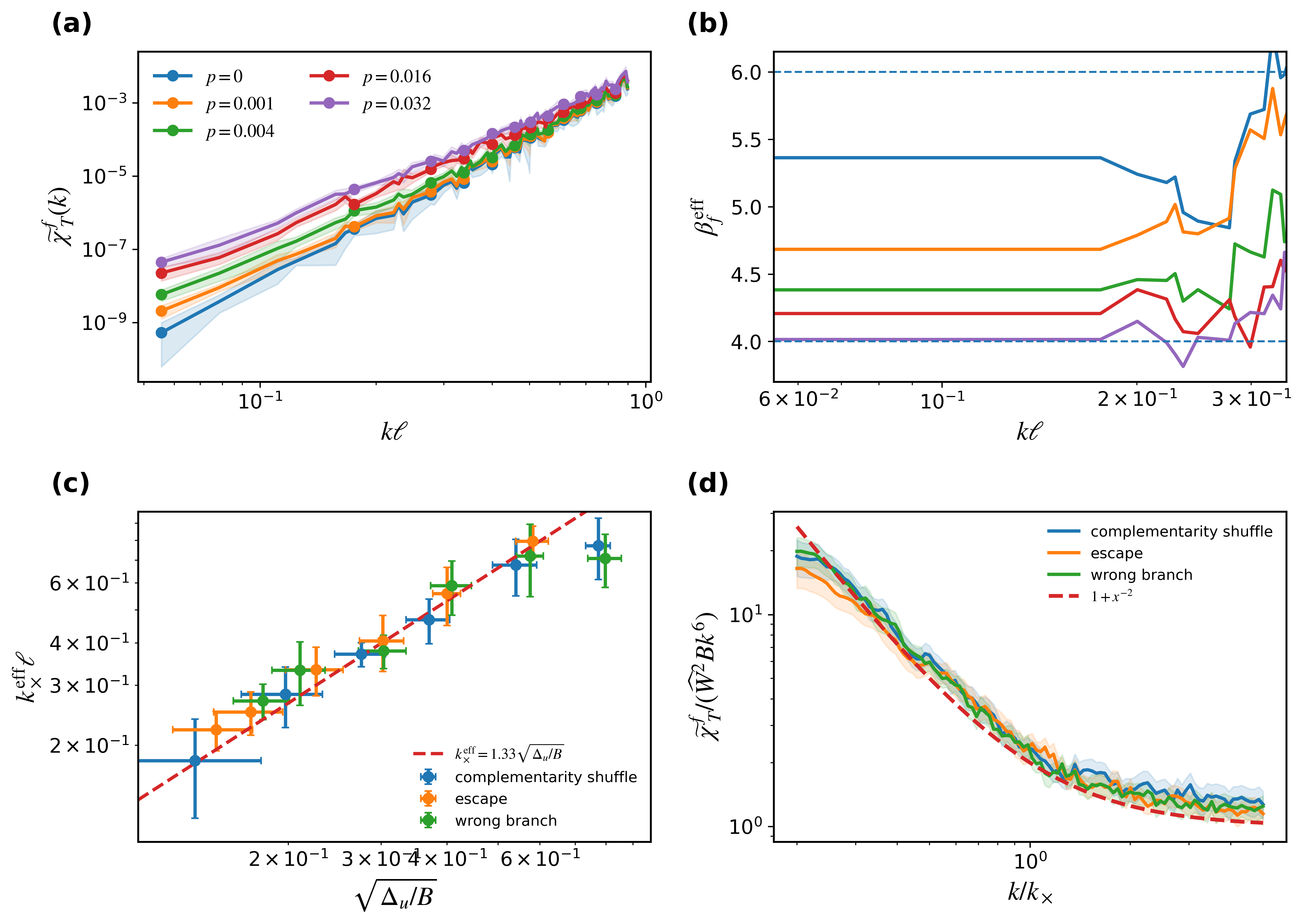}
 \caption{\textbf{Fixed-partner model: controlled unscreened-defect crossover.}
 Exact spectra, sliding effective exponents, operational crossing versus
 $\sqrt{\Deltau/B}$, and the normal-form collapse for wrong-branch,
 partial-complementarity, and box-scale-escape protocols.  Error bars are 95\%
 intervals where shown.}
 \label{fig:S11}
\end{figure}

\subsection{Three independent defect routes}
Wrong polar branches, partial moment reassignment, and box-scale separation of
complementary partners all raise $\rholoc$, raise $\Deltau$, and move the
running exponent toward 4 [Fig.~\ref{fig:S11}].  The operational
$\beta_f^{\rm eff}=5$ crossing is proportional to $\sqrt{\Deltau/B}$ with a
common finite-window coefficient $1.33$ $[1.29,1.37]$.  Wrong-branch and escape
protocols give log--log $\Deltau$--$\rholoc$ slopes $0.98$ $[0.90,1.08]$ and
$1.04$ $[0.91,1.22]$, respectively.  The complementarity protocol is less
tightly determined, $1.77$ $[0.90,2.87]$, because its weakest perturbations
remain near the baseline plateau.

\begin{figure}[t]
 \centering
 \includegraphics[width=0.62\textwidth]{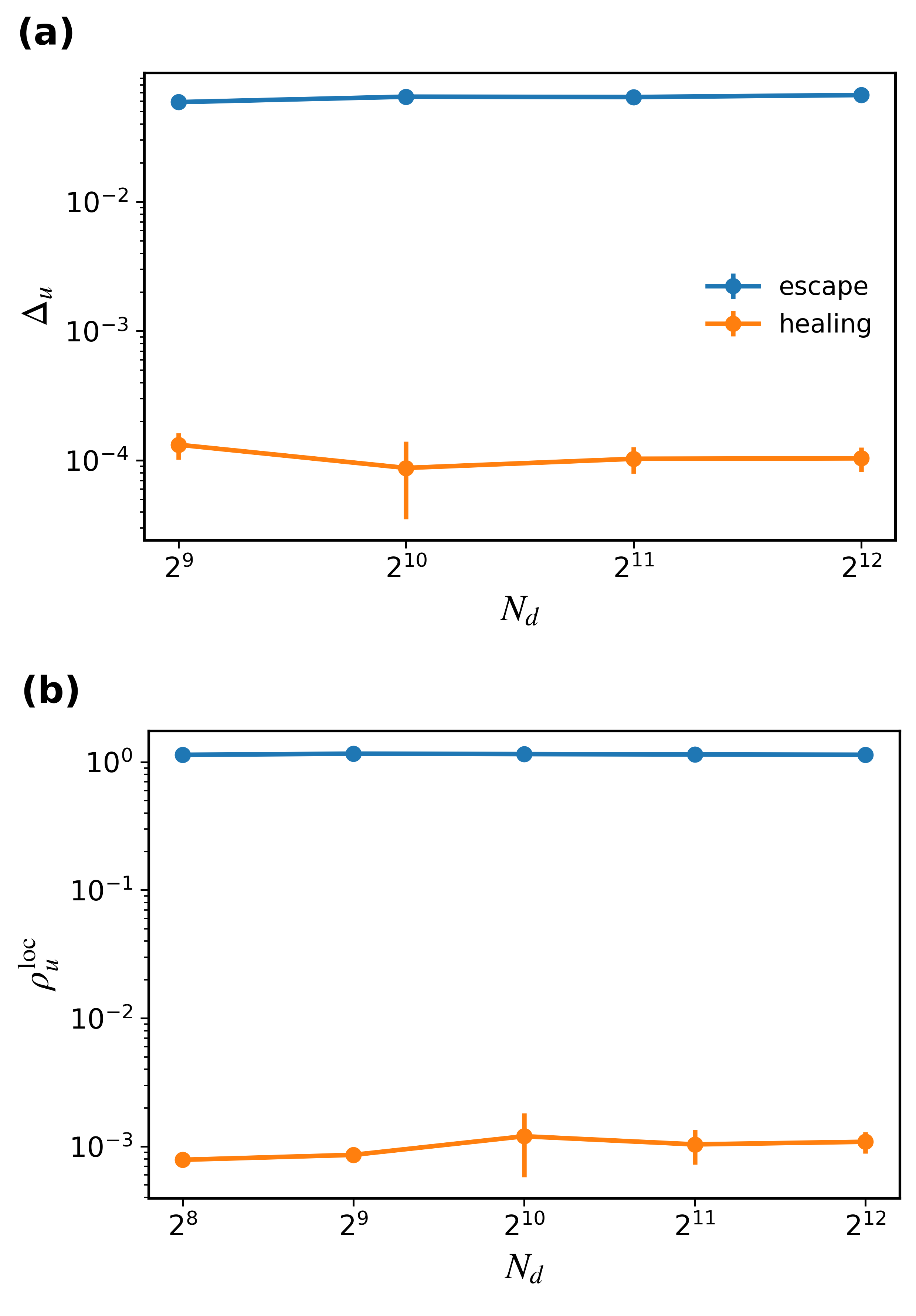}
 \caption{\textbf{Fixed-partner model: finite-size persistence.}
 Plateau and local residual in the healing and escape branches.  Error bars
 are 95\% seed-level intervals.}
 \label{fig:S12}
\end{figure}

\begin{table}[t]
\centering
\caption{Fixed-partner finite-size estimates. Mean $[95\%\ \mathrm{CI}]$ over
six seeds.}
\label{tab:legacy-size}
\begin{tabular}{ccccc}
\toprule
branch & $N_d$ & $\Deltau$ & $\rholoc$ & $\beta_f$\\
\midrule
healing & 512 & $1.32[0.92,1.73]\!\times10^{-4}$ & $0.859[0.779,0.939]\!\times10^{-3}$ & $5.51[5.38,5.63]$\\
healing & 1024 & $0.88[0.19,1.56]\!\times10^{-4}$ & $1.20[0.38,2.02]\!\times10^{-3}$ & $5.58[5.28,5.89]$\\
healing & 2048 & $1.03[0.71,1.35]\!\times10^{-4}$ & $1.04[0.62,1.45]\!\times10^{-3}$ & $5.44[5.32,5.56]$\\
healing & 4096 & $1.04[0.75,1.33]\!\times10^{-4}$ & $1.09[0.82,1.36]\!\times10^{-3}$ & $5.46[5.36,5.56]$\\
escape & 512 & $0.0591[0.0546,0.0636]$ & $1.162[1.112,1.212]$ & $4.01[3.88,4.15]$\\
escape & 1024 & $0.0651[0.0615,0.0687]$ & $1.155[1.123,1.186]$ & $3.92[3.83,4.01]$\\
escape & 2048 & $0.0647[0.0607,0.0687]$ & $1.147[1.123,1.172]$ & $3.96[3.86,4.05]$\\
escape & 4096 & $0.0670[0.0653,0.0688]$ & $1.140[1.104,1.177]$ & $3.95[3.90,3.99]$\\
\bottomrule
\end{tabular}
\end{table}

\subsection{Fixed-partner finite-size behavior}
At $N_d=4096$, the healing branch retains a small but nonzero plateau
$1.04\times10^{-4}$ $[0.75,1.33]\times10^{-4}$, whereas the escape branch has
$0.0670$ $[0.0653,0.0688]$ [Fig.~\ref{fig:S12} and
Table~\ref{tab:legacy-size}].  The corresponding local residuals are likewise
size independent.  As in the exchangeable model, the strict asymptote at any
finite residual is $k^4$ even when a broad near-$k^6$ window is visible.

\begin{figure}[t]
 \centering
 \includegraphics[width=0.96\textwidth]{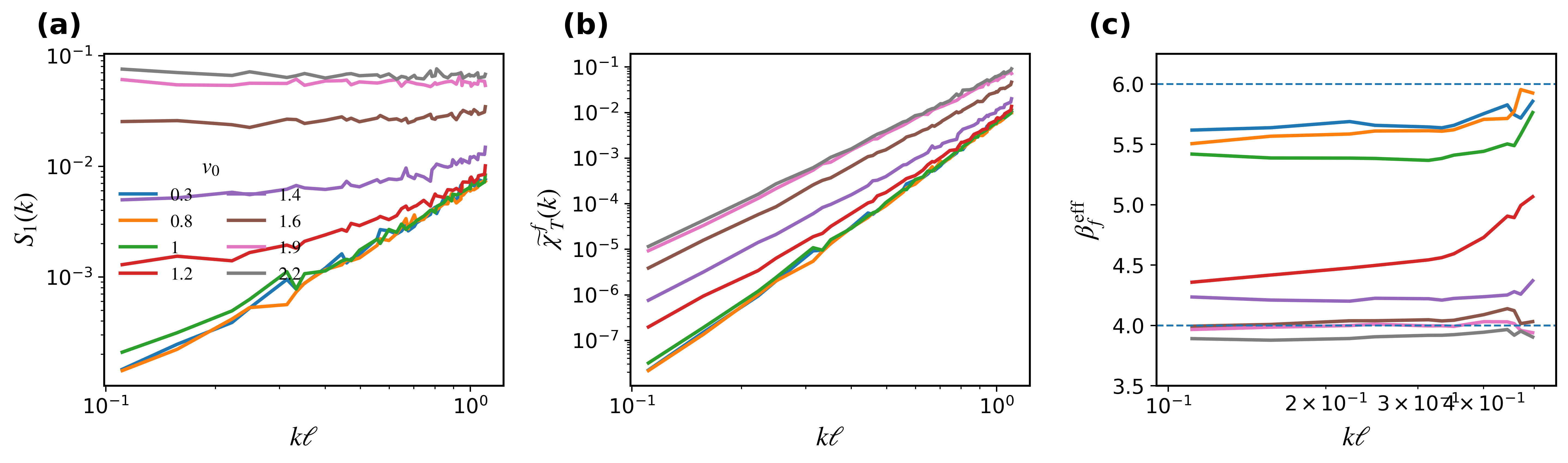}
 \caption{\textbf{Fixed-partner model: complete activity spectra.}
 First-moment spectra, force spectra, and local exponents over the full
 active-speed scan.  Curves are seed means; selected 95\% intervals are given
 in Table~\ref{tab:legacy-activity}.}
 \label{fig:S13}
\end{figure}

Figure~\ref{fig:S13} shows the continuous contraction of the screened window
as activity drives partner escape.  The evolution is a finite-system
crossover, not a sharp transition.

\begin{figure}[t]
 \centering
 \includegraphics[width=0.96\textwidth]{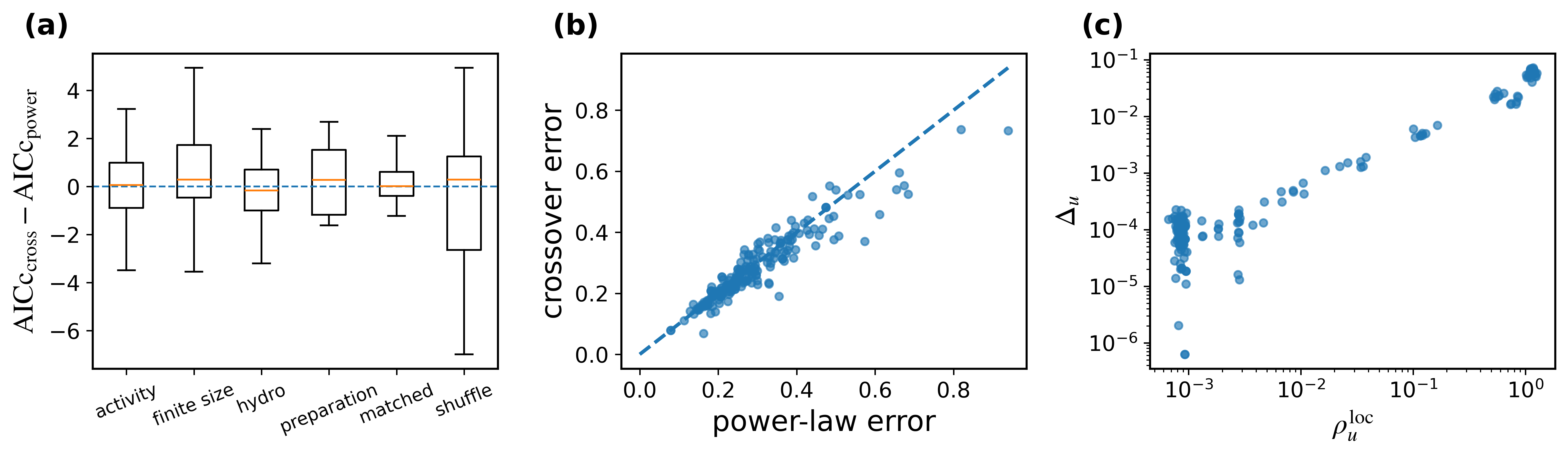}
 \caption{\textbf{Fixed-partner model: model selection and local--spectral
 correspondence.} AICc differences, held-out-shell errors, and the relation
 between $\Deltau$ and $\rholoc$.}
 \label{fig:S14}
\end{figure}

As in Model I, individual finite-window spectra do not always distinguish a
free power law from the two-term crossover decisively [Fig.~\ref{fig:S14}].
The claim is supported by analyticity, local residuals, controlled defects,
and cross-protocol scaling rather than a single information criterion.

\begin{figure}[t]
 \centering
 \includegraphics[width=0.96\textwidth]{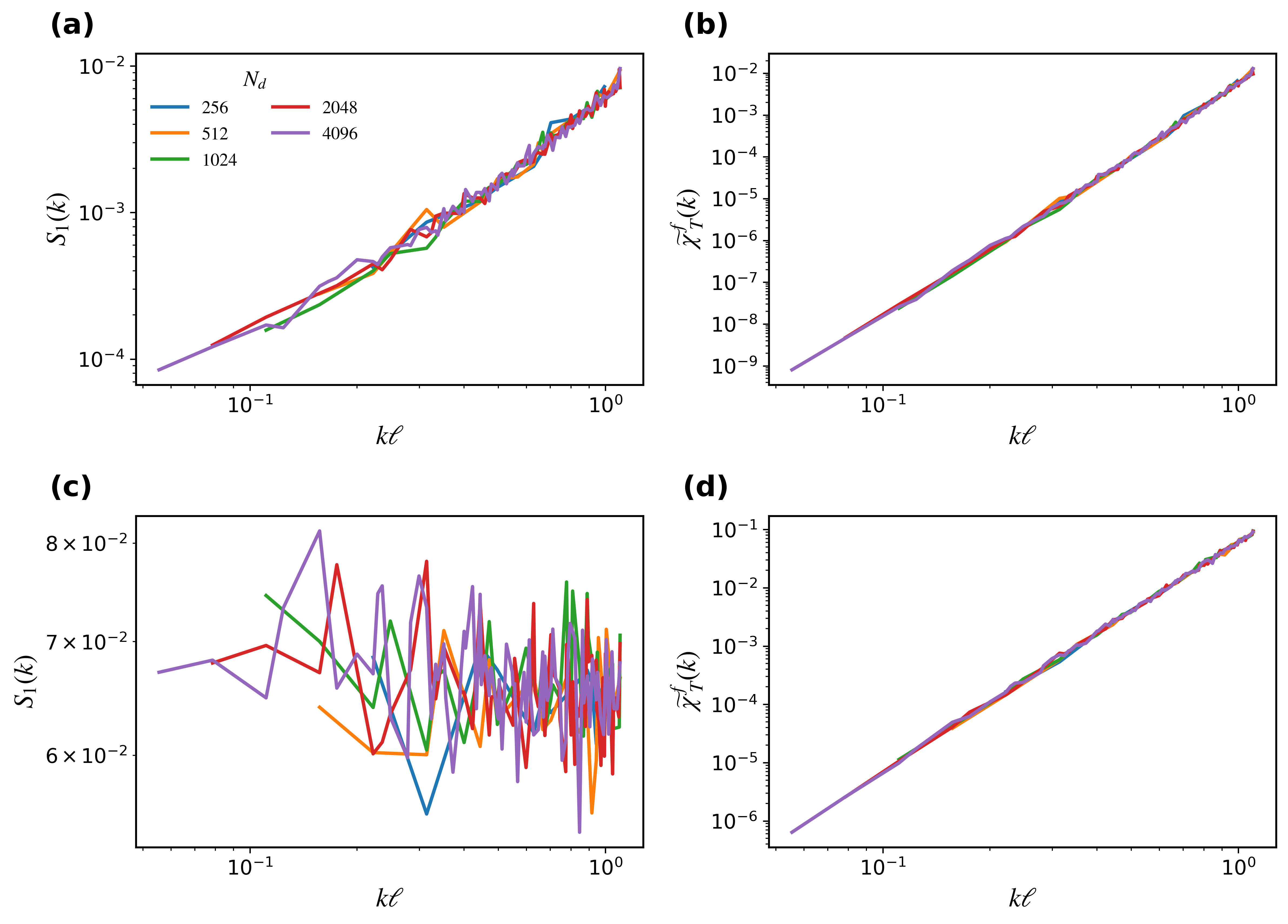}
 \caption{\textbf{Fixed-partner model: complete finite-size spectra.}
 Healing- and escape-branch first-moment and force spectra for
 $N_d=256$--4096. Curves and bands are seed means and 95\% intervals.}
 \label{fig:S15}
\end{figure}

The complete spectra in Fig.~\ref{fig:S15} show a stable $k^4$ escape branch
and a healing branch with a small plateau plus a broad near-$k^6$ window.

\begin{figure}[t]
 \centering
 \includegraphics[width=0.96\textwidth]{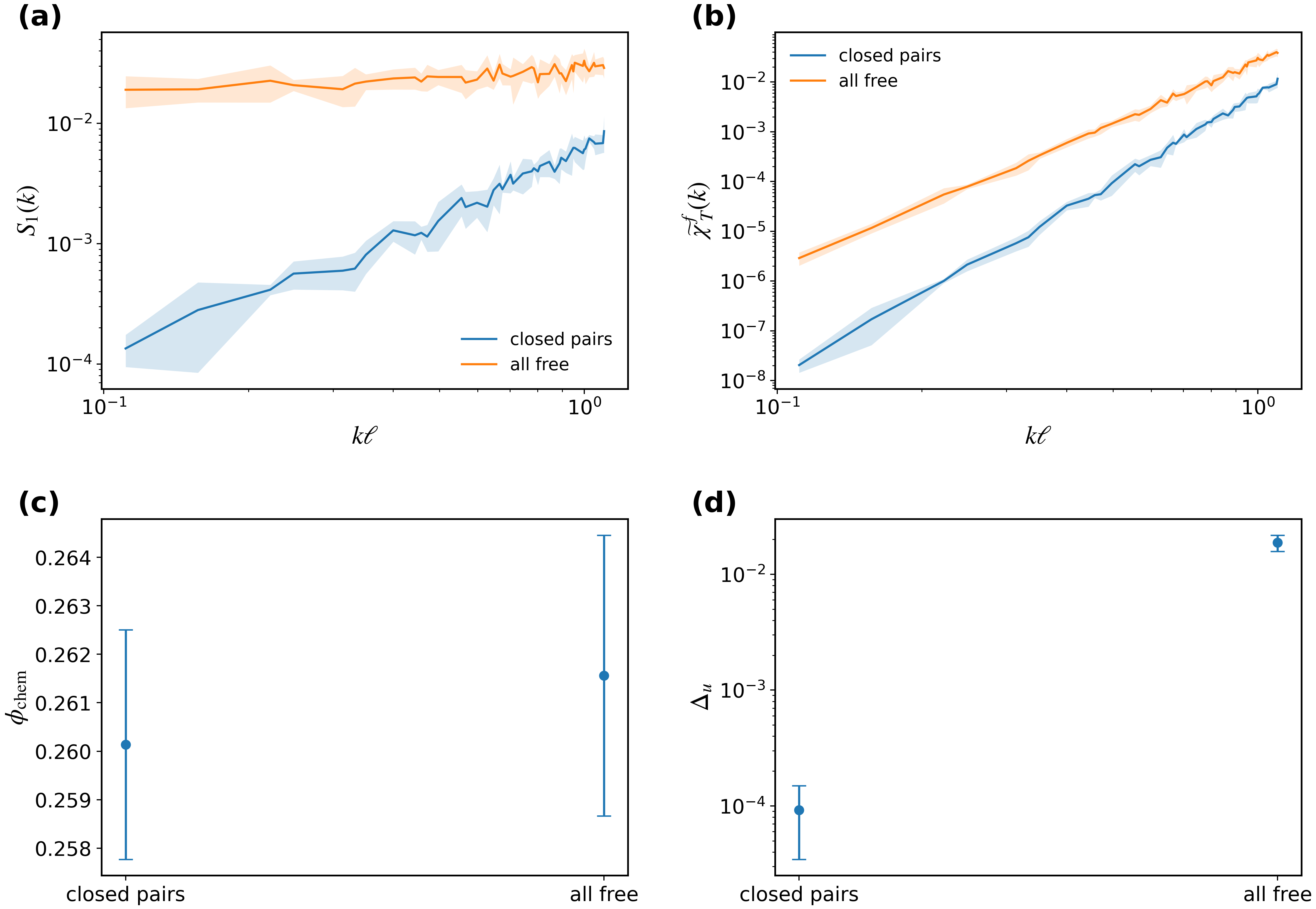}
 \caption{\textbf{Fixed-partner model: preparation dependence.}
 Similar chemical-state convergence but different hidden spectral sectors for
 closed-pair and all-free preparations. Bands and bars are 95\% seed-level
 intervals.}
 \label{fig:S16}
\end{figure}

\subsection{Preparation boundary of the fixed-partner model}
Closed-pair and all-free initial conditions reach chemical fractions $0.2601$
$[0.2578,0.2625]$ and $0.2616$ $[0.2587,0.2645]$, but their plateaus are
$9.22\times10^{-5}$ $[3.46,14.98]\times10^{-5}$ and $0.0187$
$[0.0157,0.0217]$ [Fig.~\ref{fig:S16}].  Their force exponents are $5.54$
$[5.22,5.87]$ and $4.16$ $[4.04,4.29]$.  The minimal nematic interaction does
not uniquely choose the polar $\mathbb Z_2$ branch needed for moment
cancellation.  The fixed-partner model therefore establishes maintenance,
self-healing, and destruction of a prepared complementary state, but not de
novo branch selection from every random preparation.  Model I removes this
limitation by using exchangeable opposite-sign association from random initial
conditions.

\begin{figure}[t]
 \centering
 \includegraphics[width=0.96\textwidth]{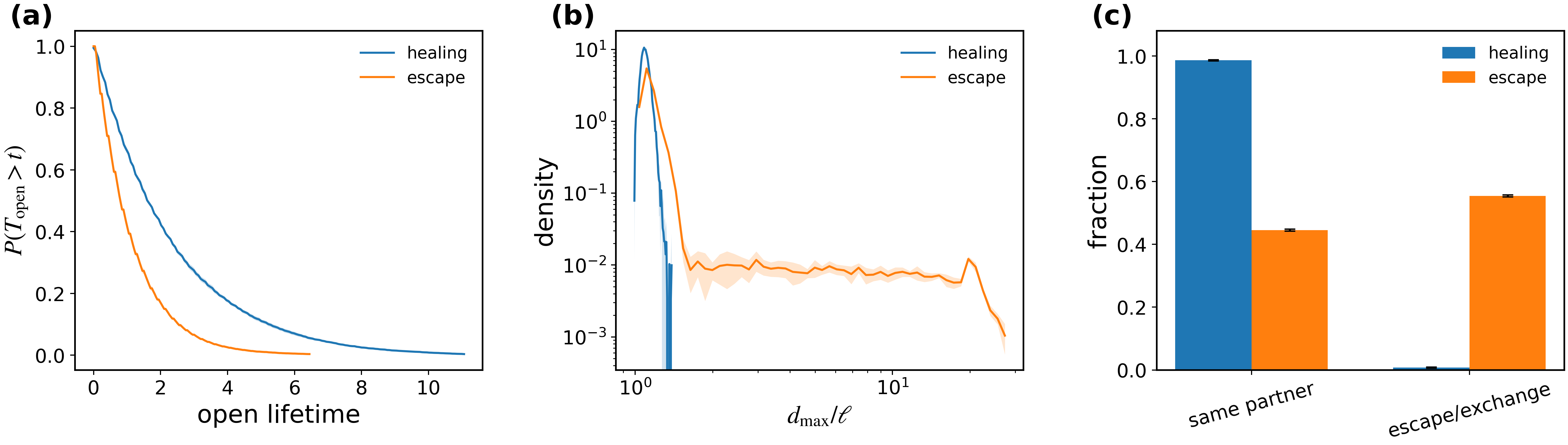}
 \caption{\textbf{Fixed-partner model: exact rupture-event distributions.}
 Open-lifetime survival, maximum separation, and outcomes in healing and
 escape regimes. Bands and bars are 95\% intervals across six seeds.}
 \label{fig:S17}
\end{figure}

\subsection{Reaction-resolved event statistics}
Events are registered inside the closed-to-open handler and store original
partner identities, exact time, maximum separation, lifetime, and outcome.
The healing same-partner return fraction is $0.9865$ $[0.9848,0.9883]$, the
median open lifetime is $1.633$ $[1.606,1.660]$, and the seed-averaged 95th
percentile of maximum separation is $1.1738\ell$ $[1.1713,1.1762]\ell$.  The
escape values are $0.4456$ $[0.4423,0.4489]$, $0.808$ $[0.787,0.830]$, and
$17.95\ell$ $[17.68,18.21]\ell$ [Fig.~\ref{fig:S17}].  The shorter lifetime
in the escape branch reflects rapid termination by escape or exchange, not
faster healing.

\begin{table}[t]
\centering
\caption{Fixed-partner exact rupture-event statistics. Mean
$[95\%\ \mathrm{CI}]$ over six seeds.}
\label{tab:legacy-events}
\begin{tabular}{ccc}
\toprule
quantity & healing & escape\\
\midrule
completed events/seed & $5424[5370,5478]$ & $6882[6809,6956]$\\
same-partner return & $0.9865[0.9848,0.9883]$ & $0.4456[0.4423,0.4489]$\\
escape/exchange & $0.00792[0.00613,0.00971]$ & $0.5544[0.5511,0.5577]$\\
median lifetime & $1.633[1.606,1.660]$ & $0.808[0.787,0.830]$\\
mean $d_{\max}/\ell$ & $1.0973[1.0968,1.0977]$ & $3.166[3.111,3.221]$\\
95th percentile $d_{\max}/\ell$ & $1.1738[1.1713,1.1762]$ & $17.95[17.68,18.21]$\\
\bottomrule
\end{tabular}
\end{table}

\begin{figure}[t]
 \centering
 \includegraphics[width=0.96\textwidth]{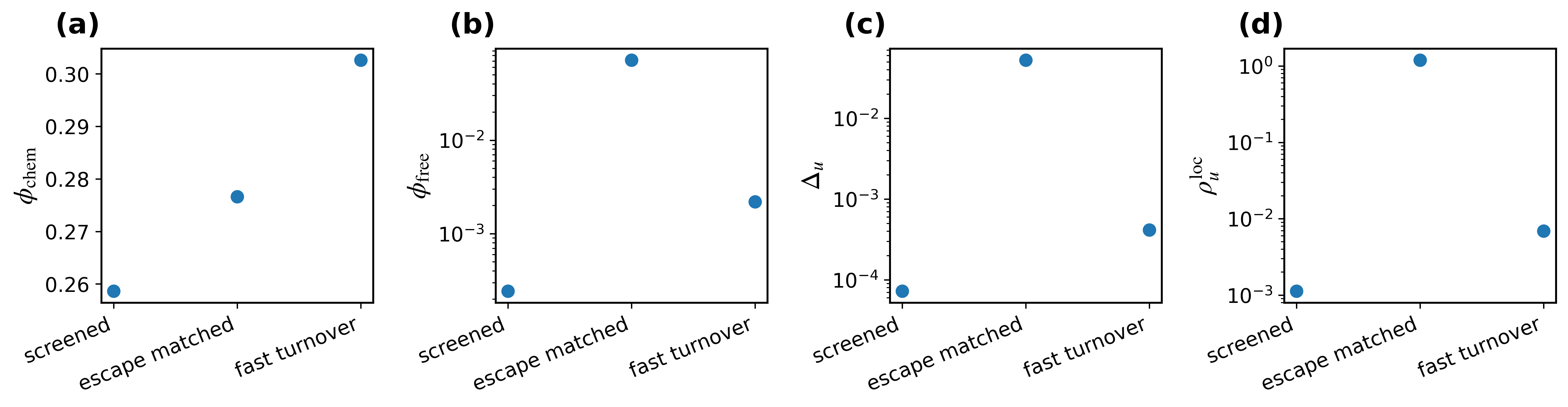}
 \caption{\textbf{Fixed-partner model: matched chemical-defect controls.}
 Chemical fraction, escaped fraction, plateau, and local residual for screened,
 escape-matched, and fast-turnover protocols. Numerical 95\% intervals are
 given in the text.}
 \label{fig:S18}
\end{figure}

\subsection{Matched-chemistry controls}
\rev{The screened state has chemical-defect fraction $0.2586$
$[0.2558,0.2615]$ and exponent $5.63$ $[5.40,5.86]$, whereas the
escape-matched state has fraction $0.2766$ $[0.2751,0.2781]$ and exponent
$3.97$ $[3.90,4.04]$.  Their plateaus differ by nearly three orders of
magnitude: $7.26\times10^{-5}$ $[1.95,12.56]\times10^{-5}$ versus $0.0527$
$[0.0480,0.0574]$ [Fig.~\ref{fig:S18}].}  A fast-turnover state can have few free fragments yet a
weaker screened window, demonstrating that residence time and spatial
complementarity, not chemical count alone, set the residual.

\begin{figure}[t]
 \centering
 \includegraphics[width=0.96\textwidth]{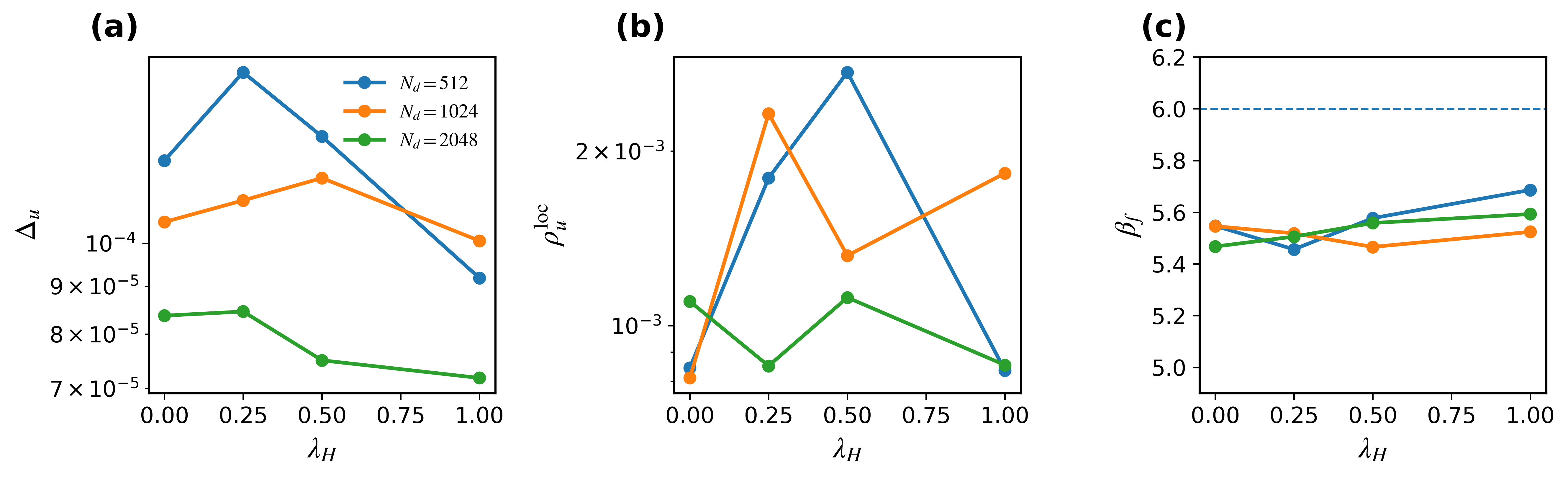}
 \caption{\textbf{Fixed-partner model: hydrodynamic continuation.}
 Plateau, local residual, and force exponent as particle--mesh Stokes coupling
 is increased from zero to one. Curves are seed means; numerical 95\%
 intervals are reported in the text.}
 \label{fig:S19}
\end{figure}

\subsection{Hydrodynamic feedback}
For $N_d=2048$, increasing $\lambda_H$ from 0 to 1 changes the exponent from
$5.467$ $[5.360,5.574]$ to $5.593$ $[5.510,5.676]$ and the plateau from
$8.37\times10^{-5}$ $[6.39,10.35]\times10^{-5}$ to
$7.18\times10^{-5}$ $[4.87,9.50]\times10^{-5}$
[Fig.~\ref{fig:S19}].  Thus the prepared screened branch persists under this
continuation.  This test does not imply that hydrodynamics selects the correct
polar branch from an all-free state.  Compressible, chiral, or odd response
operators can mix sectors and create additional leakage channels
\cite{SI@Banerjee2017,SI@Soni2019,SI@Souslov2019,SI@Scheibner2020,SI@Fruchart2021}.

\begin{figure}[t]
 \centering
 \includegraphics[width=0.96\textwidth]{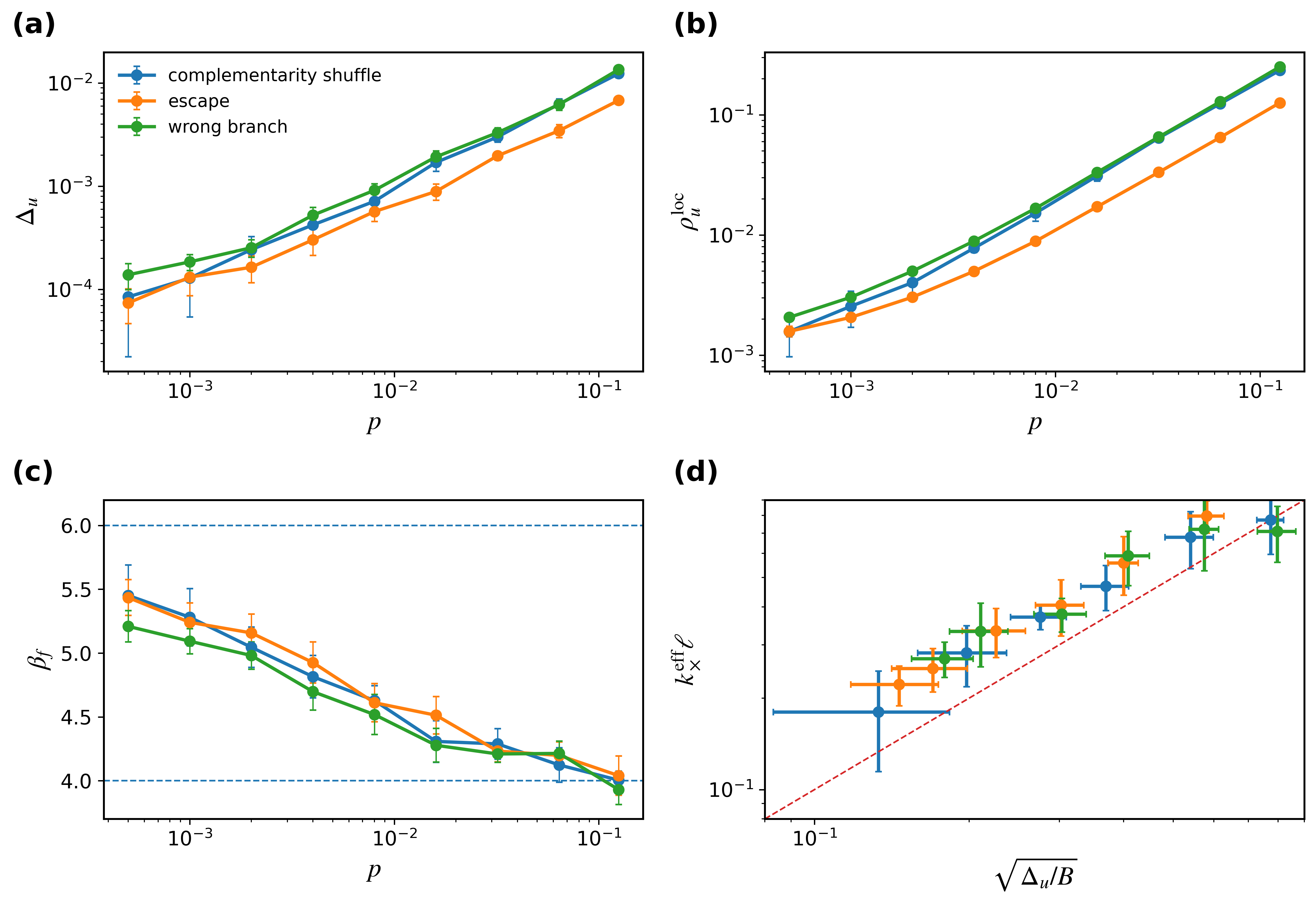}
 \caption{\textbf{Fixed-partner model: complete controlled-defect protocols.}
 Wrong-branch, partial-complementarity, and box-scale-escape scans. Error bars
 are 95\% base-seed intervals; unresolved operational crossings are omitted
 from the crossing panel but retained in the archived data.}
 \label{fig:S20}
\end{figure}

Figure~\ref{fig:S20} records the complete scans behind the three-protocol
collapse.  When $\kcross$ leaves the resolved interval, the two-term estimator
becomes poorly conditioned; those points are retained for transparency but
are not used to claim precise crossing scaling.

\section{Cross-model synthesis, generality, and limitations}
\label{sec:synthesis}

\begin{table*}[t]
\centering
\caption{Independent ingredients and common consequences of the two models.}
\label{tab:model-comparison}

\begin{minipage}[t]{0.47\textwidth}
\raggedright
\textbf{Model I: exchangeable carriers.}
The carriers have no prescribed partners: each carrier can bind to any
compatible neighbor while maintaining unit valence. The system is
initialized from a random, unbound configuration. Its microscopic state
is specified by the carrier position, polar axis, sign, and instantaneous
bond index. The principal control parameter is the turnover rate, and the
independent local diagnostic is the density of unpaired moments. This
model therefore directly demonstrates random self-assembly accompanied
by complete partner renewal.
\end{minipage}
\hfill
\begin{minipage}[t]{0.47\textwidth}
\raggedright
\textbf{Model II: internally fluctuating molecules.}
The molecular constituents carry complementary partner labels and
transition among closed, open, and free states. They may be initialized
either through a prepared complementary protocol or from an all-free
state. Each molecule is described by its center, nematic axis, sign,
length, and reaction state. The relevant controls include activity,
escape processes, branch defects, and complementarity. Its independent
local diagnostic combines the pair residual with the free-moment
density. This formulation provides event-resolved causality and three
independently controlled routes for generating defects.
\end{minipage}

\medskip
\hrule
\medskip

\begin{minipage}{0.94\textwidth}
\raggedright
\textbf{Common long-wavelength consequences.}
Despite their distinct microscopic ingredients, both models exhibit the
same moment spectrum,
\begin{equation}
    \Sone(k)=\Deltau+B k^2,
\end{equation}
and consequently the same force spectrum,
\begin{equation}
    \Sf(k)=\Deltau k^4+B k^6.
\end{equation}
They therefore generate the same velocity-spectrum crossover, while
providing independent microscopic realizations of the underlying
long-wavelength mechanism.
\end{minipage}

\end{table*}

The common result is not that every active material is hyperuniform.  It is a
conditional infrared statement: if the response-selected first moments are
organized into finite locally neutral clusters, their contribution is
analytic and begins at $k^2$; if a finite density of unpaired or imperfect
clusters remains, it contributes a plateau.  Cluster membership, lifetime,
reaction chemistry, and internal coordinates determine the amplitudes
$\Deltau$ and $B$, but not the two leading powers.  Model I shows that permanent
partner identity and prepared branch labels are unnecessary.  Model II shows
that chemical defects, escaped fragments, wrong branches, and complementarity
loss all act through the same residual variable.

The conclusions have four boundaries.  First, the observed noninteger window
exponents are crossovers, not universal fractional asymptotes.  Second, the
theory assumes finite cluster size and finite-range correlations; critical
cluster distributions could generate nonanalytic corrections.  Third, the
new model uses a minimal active first-moment carrier rather than a chemically
specific experimental species.  Its value is to establish a general mechanism
under reversible self-assembly, not to fit one material parameter by parameter.
Fourth, the response analysis is incompressible and parity even.  Chiral,
compressible, or odd media may mix additional sectors.

Within these boundaries, the two independent models support a sharp physical
conclusion: exchange does not destroy multipole inheritance, but any finite
unpaired residual makes the strict infrared limit singular.  The appropriate
order parameter is therefore the locally unscreened moment density and its
spectral plateau, while the experimentally relevant measure of quietness is
the screening length over which the higher-order response survives.

\section{Relation to active-matter and hyperuniformity literature}
The response-selected signed-moment mechanism differs from scalar density
hyperuniformity in driven suspensions, circle swimmers, chiral active fluids,
and active field theories
\cite{SI@Corte2008,SI@HexnerChaikinLevine2017,SI@Wilken2020,SI@LeiCiamarraNi2019,SI@LeiNi2019,SI@Huang2021,SI@ZhangSnezhko2022,SI@BackofenVoigt2024,SI@ZhengKlattLowen2024,SI@MaireChaix2025,SI@Maire2026ActiveNucleation}.
It also differs from phase separation, active glassiness, caging, crystalline
order, and absorbing transitions
\cite{SI@TailleurCates2008,SI@CatesTailleur2015,SI@FodorCates2016,SI@Tjhung2018,SI@NessCates2020,SI@CatesNardini2025,SI@Berthier2014,SI@FlennerBerthierSzamel2016,SI@BerthierFlennerSzamel2019,SI@Charbonneau2025ActiveCaging,SI@MorseCorwinManning2021,SI@PraetoriusVoigt2018,SI@Reichhardt2023EPL}.
The defining feature here is that the physical Green function selects a signed
multipole sector whose chemical, geometric, and spectral defect measures are
not interchangeable.

\clearpage

\begingroup
\makeatletter
\global\let\@FMN@list\@empty
\let\label\@gobble
\makeatother

\endgroup

\end{document}